\documentclass[reprint,nofootinbib,amsmath,amssymb,aps,floatfix]{revtex4-2}
\usepackage{graphicx}
\usepackage{dcolumn}
\usepackage[multiple]{footmisc}
\usepackage{appendix}
\usepackage{color}
\usepackage{multirow}
\usepackage{subfigure}
\usepackage{hyperref,verbatim}

\usepackage{float}

\begin{document}

\title{Radiative properties of thin accretion disks around rotating short-hairy black holes}
\author{Shou-Qi Liu}%\textsuperscript{1}}
\author{Jia-Hui Huang}%\textsuperscript{1,2}} 
\email{huangjh@m.scnu.edu.cn}
\date{\today}

\affiliation{%$^{1}$
Key Laboratory of Atomic and Subatomic Structure and Quantum Control (Ministry of Education), Guangdong Basic Research Center of Excellence for Structure and Fundamental Interactions of Matter, School of Physics, South China Normal University, Guangzhou 510006, China}
\affiliation{%$^{2}$
Guangdong Provincial Key Laboratory of Quantum Engineering and Quantum Materials, Guangdong-Hong Kong Joint Laboratory of Quantum Matter, South China Normal University, Guangzhou 510006, China}

\begin{abstract}
The radiative properties of the accretion disks around rotating short-hairy black holes are studied. The geodesic motion of massive particles in the equatorial planes of short-hairy black holes and impact of the rotation parameter $a$, short-hair parameter $Q$ and power parameter $k$ on the geodesic motion is obtained. Then, we consider thin accretion disks around the short-hairy black holes. The radiative flux densities, effective temperatures, differential and spectral luminosity of the disks are investigated for various specific parameters. The impact of parameters $a,Q,k$ on these properties is also discussed. 
\end{abstract}
\maketitle

\section{Introduction}
The existence of black holes are predicted by Einstein's general relativity. Significant progresses have been made for observational evidence of black holes in recent years. 
The gravitational wave signals produced by merger of black holes have been detected by LIGO and Virgo collaboration since 2015 \cite{LIGOScientific:2016aoc}.   
The image of the supermassive black hole $\text{M}87^*$ was provided by Event Horizon Telescope in 2019 \cite{M87-EHT-1} and the image of the supermassive black hole Sagittarius $A^*$ in our Galactic Center was also released in 2022\cite{SgrA-EHT-1}. 
These findings confirm the existence of astrophysical black holes, which are modeled by the Kerr metric\cite{Kerr:1963ud,LIGOScientific:2016aoc}.

The Kerr black hole is the unique stationary, axially symmetric and asymptotically flat vacuum solution of Einstein's equation of gravitational field, which is guaranteed by the no-hair theorems or uniqueness theorems \cite{Ruffini:1971bza,Robinson:1975bv,Hawking:1971vc,Carter:1971zc,Israel:1967wq,Mazur:1982db,Gurlebeck:2015xpa,Bekenstein1995,Ghosh:2023kge,Ghosh:2025igz}. However, counterexamples and violation of the no-hair theorems have also been investigated in literature \cite{Chrusciel:2012jk,Herdeiro:2015waa} , which usually occurs when gravity is coupled with matter fields, such as gauge fields and scalar fields. These black holes are usually called hairy black holes, which include non-Abelian black holes \cite{Bizon:1990sr},
black holes with Skyrme hair \cite{Droz:1991cx}, Higgs hair \cite{Breitenlohner:1991aa,Greene:1992fw,Hartmann:2001ic}, massive graviton hair \cite{Brito:2013xaa} or dilaton hair \cite{Lavrelashvili:1992ia,Kanti:1995vq,Lee:2018zym}. More recent examples include 
scalar hairy black holes from coupling between scalar functions and the Gauss-Bonnet term \cite{Sotiriou:2014pfa,Antoniou:2017acq,Silva:2017uqg,Doneva:2017bvd,Doneva:2021tvn,Macedo2019sem,Kuan:2021lol,Dima:2020yac,Berti:2020kgk,Herdeiro:2020wei,Cunha:2019dwb,Hod:2020jjy}, hairy rotating black holes induced by superradiant instability \cite{Herdeiro:2014goa,Herdeiro:2016tmi,Santos:2020pmh}, hairy black holes supported by nonlinear scalar potentials \cite{Corichi:2005pa,Feng:2013tza,Chew:2022enh,Chew:2024rin,Chew:2023olq,Chew:2024evh}, charged hairy black holes \cite{Herdeiro:2018wub,Fernandes:2020gay,Brihaye:2021mqk,Hong:2020miv,Hong:2019mcj,Herdeiro:2020xmb,Zhang:2021nnn,Zou:2019bpt}, and black holes with Proca \cite{Herdeiro:2016tmi,Santos:2020pmh,Barton:2021wfj} and electroweak hairs \cite{Gervalle:2024yxj}, etc.\cite{Doneva:2022ewd}. 

The astrophysical black holes usually live in an non-vacuum environment and couple with surrounding matter. Thus, hairy black holes should provide a better fit for astrophysical black holes. In Ref.\cite{Brown:1997jv}, the authors considered the coupling of gravity to an anisotropic fluid and obtained a class of static, spherically symmetric black holes with  short hair, which violates the argument that black hole has no short hair \cite{Nunez:1996xv}. Furthermore, these black hole solutions were extended to rotating ones and the impact of the short-hair parameter on the black hole shadow was also studied \cite{Tang:2022uwi}. Recently, test of the weak cosmic censorship conjecture \cite{Zhao:2024qzg} and the strong gravitational lensing effects \cite{Zhao:2024hep} have also been explored in the rotating short-hairy black hole model. It will be interesting to study other properties of the rotating short-hairy black holes. For example, exploring the properties of the accretion disk around the rotating short-hairy black hole can help us further understand the near-horizon characteristics of the black hole and test the gravitational theories.

In this paper, we consider the properties of the emission spectrum of the accretion disk around the rotating short-hairy black hole \cite{Tang:2022uwi}. 
We consider the geometrically thin disk models \cite{Shakura:1972te,Page:1974he,Thorne:1974ve}. These thin disk models are relatively simple, the mass accretion rate is supposed to be constant in time and is independent of the radius of the disk, and most of the gravitational energy is released by radiation which generates the luminosity of the disk. For typical accretion disks, the mass of the disk is completely negligible \cite{Bambi:2014koa}.  
The electromagnetic spectrum of the accretion disk around a black hole could be used to determine the characteristics of the central black hole and the underlying gravity theories \cite{Kong:2014wha,Pun:2008ua,Harko:2009rp,Harko:2009kj,Harko:2010ua,Chakraborty:2014eha,Pun:2008ae,Li:2004aq,
Narayan:1993bd,Bambi:2017khi,Heydari-Fard:2010agr,Dyadina:2010agr,Heydari-Fard:2023kgf,Donmez:2024lfi,Boshkayev:2023fft,Perez:2017spz,Karimov:2018whx,
Jiang:2024njc,Feng:2024iqj,Heydari-Fard:2022xhr,Heydari-Fard:2021ljh,Liu:2021yev,Heydari-Fard:2020iiu,Heydari-Fard:2020ugv}.

The organization of this paper is as follows. In Sec.II, we introduce the metric of the rotating short-hairy black hole and discuss the geodesic motion of massive particles in the equatorial plane for various choices of the rotation parameter $a$, short hair parameter $Q$ and power parameter $k$. In Sec.III, the radiative properties of a thin accretion disk around the short-hairy black hole are investigated. The impact of parameters $a,Q,k$ on these properties is also numerically studied. Sec.IV is devoted to the summary.

\section{Rotating short-hairy black holes and equatorial circular orbits}
The rotating short-hairy black holes we are interested in are proposed in \cite{Tang:2022uwi}, which are obtained from spherically symmetric ones \cite{Brown:1997jv} by using the 
Newman-Janis algorithm. The line element of the rotating short-hairy black hole is
\begin{align}
ds^{2}=&-(1-\frac{2M r-\frac{Q^{2k}}{r^{2k-2}}}{\rho^{2}})d t^{2}+\frac{\rho^{2}}{\Delta}d r^{2}\nonumber\\
 &-\frac{2a\sin^{2}\theta(2M r-\frac{Q^{2k}}{r^{2k-2}})}{\rho^{2}}dt d\phi\nonumber\\
 &+\rho^{2}d\theta^{2}+\frac{\Sigma\sin^{2}\theta}{\rho^{2}}d\phi^{2},
\end{align}
where
\begin{equation}
\begin{aligned}
&\rho^{2}=r^{2}+a^{2}\cos^{2}\theta,\\
&\Sigma=(r^{2}+a^{2})^{2}-a^{2}\Delta,\\
&\Delta=r^{2}-2M r+a^{2}+\frac{Q^{2k}}{r^{2k-2}}.
\end{aligned}
\end{equation}
Here $M$ is the black hole mass, $a$ is the rotation parameter of the black hole, and $Q$ represents the short-hair parameter. $k$ is a power parameter in the equation of state of the anisotropic fluid \cite{Brown:1997jv}. When $k=1$, the metric degenerates into the Kerr-Newman black hole. When $k>1$, $Q$ qualifies as a short hair. When $k=2,~ a=0$, the metric degenerates into the quantum-corrected black hole proposed in \cite{Lewandowski:2022zce}. 

\begin{figure}[]
%\vspace{-0.8cm} %调整图片与上文的垂直距离
\centering
    \subfigure{
    \begin{minipage}{.45\linewidth}
    \centering
    \includegraphics[width=\linewidth]{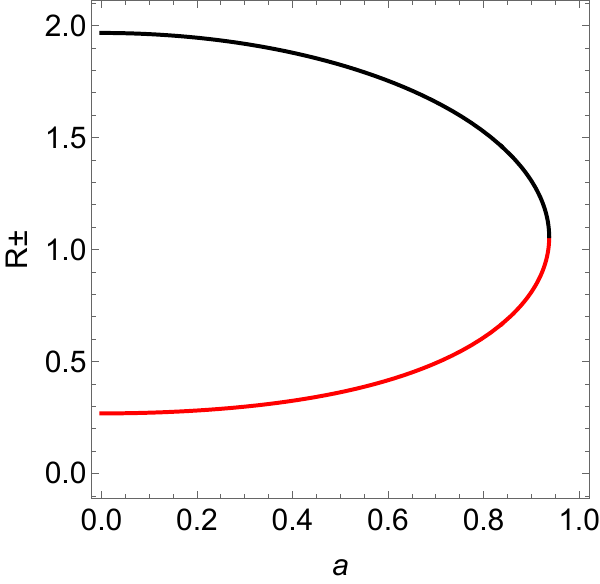 }
     \end{minipage}}     
   \subfigure{ \begin{minipage}{.45\linewidth}
   \centering
    \includegraphics[width=\linewidth]{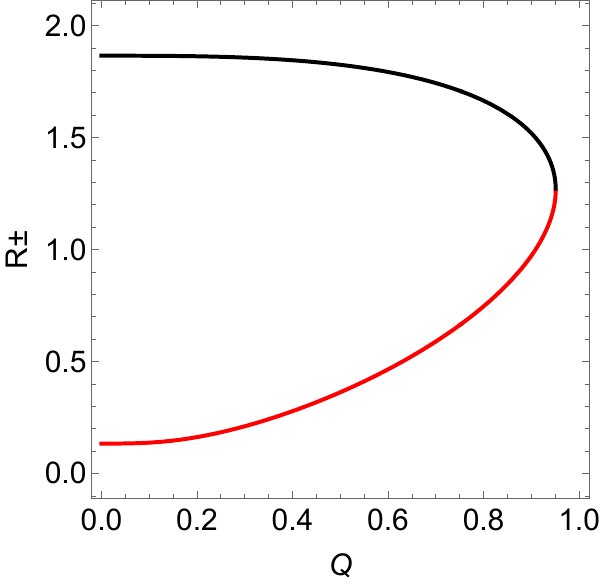}
   \end{minipage}}
   
     \subfigure{
    \begin{minipage}{.45\linewidth}
    \centering
    \includegraphics[width=\linewidth]{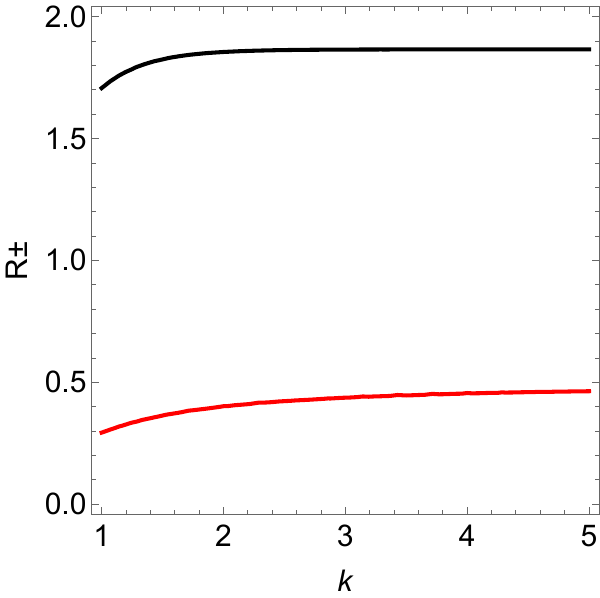}
     \end{minipage}}
       \caption{Impact of parameters $a,Q,k$ on the radii of the inner horizon (red) and outer horizon (black). The parameters are chosen as: $Q=0.5, k=1.5$ for the top left panel, $a=0.5, k=1.5$ for the top right panel and $a=0.5, Q=0.5$ for the bottom panel.}
    \label{fig:enter-label1}
\end{figure}

The horizons of the short-hairy black hole are determined by the equation $\Delta=0$. This equation can be solved numerically and we find two horizons, the inner horizon ($R_-$) and outer horizon ($R_+$). The typical relationships between the radii of the horizons and the parameters $a, Q, k$ are shown in Fig.\ref{fig:enter-label1}.
 It is easy to see that as $a$ or $Q$ increases, the radius of the outer horizon decreases while that of the inner horizon increases, and they meet at an extremal point. And the parameter $k$ has no remarkable impact on the radii of the horizons.

Next, we consider the particle motion in the equatorial plane. We scale quantities with the black hole mass $M$ and take $M=1$ hereafter.
The coordinate $\theta$ is fixed as $\frac{\pi}{2}$, then the components of the metric in the equatorial plane are 
\begin{align}
&g_{tt}=-1 +\frac{2}{r}-\frac{Q^{2k}}{r^{2k}},\\
&g_{r r}=\frac{r^{2}}{r^{2}-2r+a^{2}+{Q^{2k}}{r^{2-2\,{k}}}},\\
&g_{t \phi}=-\frac{{a} (2r-{ Q}^{2k} r^{2-2k})}{{ r}^{2}},\\
&g_{\theta\theta}=r^2,\\
&g_{\phi\phi}=\frac{(r^{2}+a^{2})^{2}-{ a}^{2}(r^{2}-2r+a^{2}+{Q}^{2k}r^{2-2k})}{r^{2}}.
\end{align}
With the above components, the Lagrangian of a test particle moving in the equatorial plane is
\begin{equation}
\begin{aligned}
\mathcal{L}= \frac{1}{2}g_{\mu\nu}\dot{x}^{\mu} \dot{x}^{\nu}=\frac{1}{2}(\dot{r}^{2}g_{rr}+\dot{t}^{2}g_{tt}+2\dot{t}\dot{\phi}g_{t\phi}+\dot{\phi}^{2}g_{\phi\phi}).
\end{aligned}
\end{equation}
The dot means derivative with respect to the proper time of the particle. 
There are two conserved quantities for the test particle, namely, the conserved specific energy $\tilde{E}$ and conserved specific angular momentum $\tilde{L}$, which are defined as
\begin{align}
 \tilde{E}&=-\mathcal{L}_{,\dot{t}}=-\dot{t}~{g_{t t}}-\dot{\phi}~{g_{t \phi}},\\   
  \tilde{L}&=\mathcal{L}_{,\dot{\phi}}=\dot{t}~g_{t\phi}~+\dot{\phi}~g_{\phi\phi}.  
\end{align}
By solving the above two equations, we can get
\begin{align}
\dot{t}&=-\frac{\tilde{L}g_{t \phi}+\tilde{E} g_{\phi \phi}}{g_{t \phi}^2-g_{tt} g_{\phi \phi}}\nonumber\\ 
&=\left(r^{2k}~(\tilde{E}r^3+a^2(2+r)\tilde{E}-2a\tilde{L})\right.\nonumber\\ 
&\left.~~~-a Q^{2k}(a\tilde{E}-\tilde{L})r\right)A^{-1}, \label{equation:enter-labe113}\\
\dot{\phi}&=-\frac{\tilde{L} g_{tt}+\tilde{E}g_{t \phi}}{g_{t \phi}^2-g_{tt} g_{\phi \phi}}\nonumber\\
 &=\left(r^{2k}(\tilde{L}(r-2)+2a\tilde{E})+Q^{2k}(\tilde{L}-a\tilde{E})r\right)A^{-1},
    \label{equation:enter-labe114}
\end{align}
where
\begin{equation}
   A= Q^{2k}r^3+r^{1+2k}(r^2-2r+a^2).
\end{equation}
For a massive particle, its four velocity also satisfies the following normalization condition
\begin{equation}
g_{\mu\nu}\dot{x}^{\mu}\dot{x}^{\nu}=-1.
 \label{equation:enter-label15}
\end{equation}

%fig. 2
 \begin{figure*}[]
    \centering
\subfigure{
    \begin{minipage}{.31\linewidth}
    \centering
    \includegraphics[width=\linewidth]{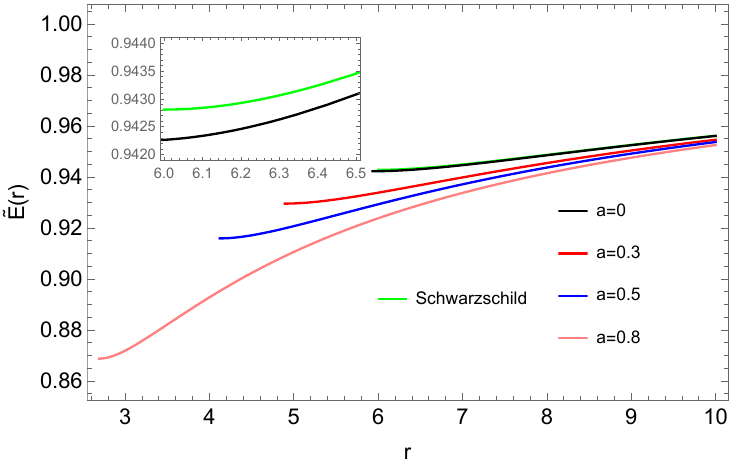}
     \end{minipage}}      
\subfigure{ \begin{minipage}{.31\linewidth}
    \centering
    \includegraphics[width=\linewidth]{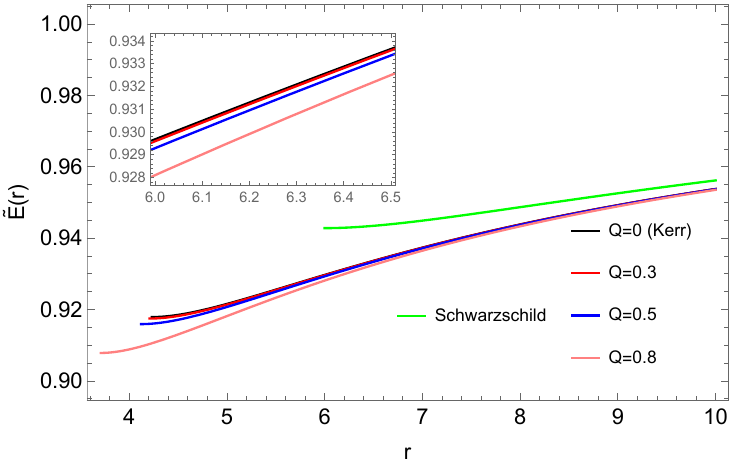}
   \end{minipage}}   
\subfigure{
    \begin{minipage}{.31\linewidth}
    \centering
    \includegraphics[width=\linewidth]{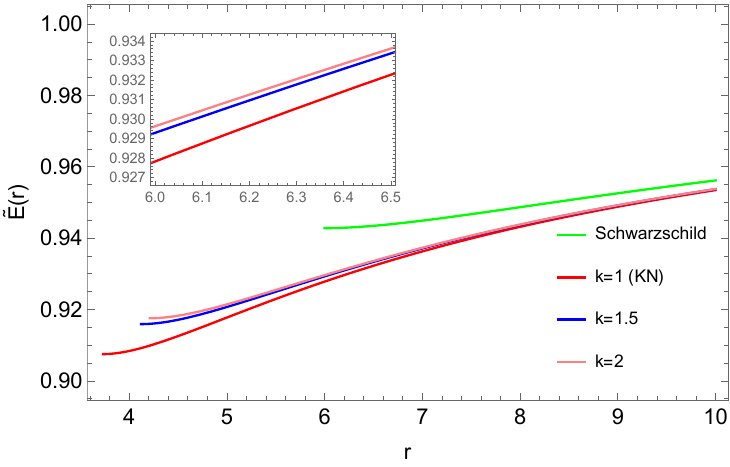}
     \end{minipage}}
       \caption{The specific energy of particles on different circular orbits for various parameters $a,Q,k$. The parameters are chosen as: $Q=0.5, k=1.5$ for the left panel, $a=0.5, k=1.5$ for the middle panel and $a=0.5, Q=0.5$ for the right panel.}
    \label{fig:energy}
\end{figure*}
%fig. 3
\begin{figure*}[]
    \centering
\subfigure{
    \begin{minipage}{.31\linewidth}
    \centering
    \includegraphics[width=\linewidth]{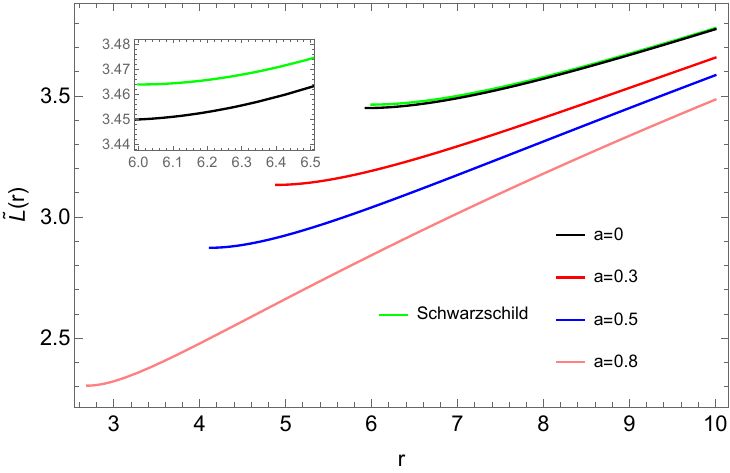}
     \end{minipage}}     
\subfigure{ \begin{minipage}{.31\linewidth}
    \centering
    \includegraphics[width=\linewidth]{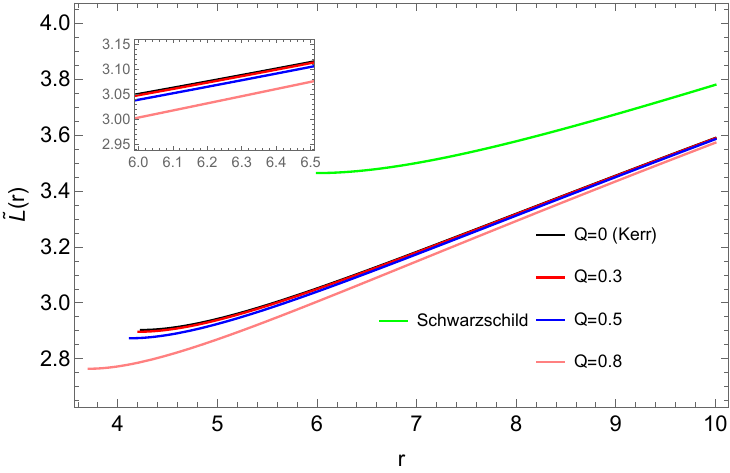}
   \end{minipage}}
 \subfigure{
    \begin{minipage}{.31\linewidth}
    \centering
    \includegraphics[width=\linewidth]{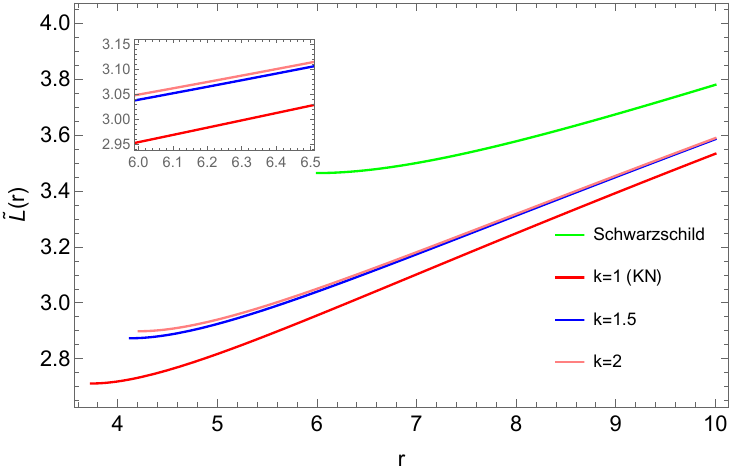}
     \end{minipage}}
       \caption{The specific angular momenta of particles on different circular orbits for various parameters $a,Q,k$. The parameters are chosen as: $Q=0.5, k=1.5$ for the left panel, $a=0.5, k=1.5$ for the middle panel and $a=0.5, Q=0.5$ for the right panel.}
    \label{fig:angm}
\end{figure*}

Plugging equations \eqref{equation:enter-labe113} and \eqref{equation:enter-labe114} into the above equation, we have
\begin{equation}\label{}
g_{rr}\dot{r}^{2}+\left(1-\frac{{ \tilde{E}}^{2}g_{\phi\phi}+2 \tilde{E} \tilde{L}g_{t\phi}+ \tilde{L}^{2}g_{t t}}{g_{t\phi}^{2}-g_{t t}g_{\phi\phi}}\right)=0
\end{equation}
The effective potential is
\begin{equation}
\begin{split}
   &V_{\mathrm{eff}}=1-\frac{{ \tilde{E}}^{2}g_{\phi\phi}+2 \tilde{E} \tilde{L}g_{t\phi}+ \tilde{L}^{2}g_{t t}}{g_{t\phi}^{2}-g_{t t}g_{\phi\phi}}\\
    &=(r^{2k}(-a^2r-(-2+r)(\tilde{L}^2+r^2)-4a\tilde{L}\tilde{E}\\
    &~~~~+(r^3+a^2 r+2a^2)\tilde{E}^2)-Q^{2k}r(r^2+(\tilde{L}-a\tilde{E})^2))\\ 
    &~~~~(Q^{2k}r^3+r^{1+2k}(r^2-2r+a^2))^{-1}
  \end{split}
\end{equation}
By solving
\begin{equation}
V_{\mathrm{e f f}}(r)=0,~ V_{\mathrm{e f f}}^{\prime}(r)=0,
\end{equation}
we can find the specific energy and the specific angular momenta of the particles moving on circular orbits with radius $r$,
\begin{align}
   \tilde{E}&=-\frac{g_{t t}+g_{t\phi}\Omega}{\sqrt{-g_{t t}-2g_{t\phi}\Omega-g_{\phi\phi}\Omega^{2}}},\\
    \tilde{L}&=\frac{g_{t\phi}+g_{\phi\phi}\Omega}{\sqrt{-g_{t t}-2g_{t\phi}\Omega-g_{\phi\phi}\Omega^{2}}},
\end{align}
where $\Omega=\frac{d\phi}{dt}=\frac{\dot{\phi}}{\dot{t}}$ is the angular velocity of the particle. For particles on circular orbits, 
the expressions for the angular velocities are
\begin{align}
 & \Omega_{\pm}=\frac{-g_{t\phi,r}\pm\sqrt{(g_{t\phi,r})^{2}-g_{t t,r}g_{\phi\phi,r}}}{g_{\phi\phi,r}}\\
 &=\frac{a k Q^{2k}r-a r^{2k}\pm r^{2+2k}\sqrt{\frac1r-k Q^{2k}r^{-2k}}}{a^2kQ^{2k}r+r^{2k}(-a^2+r^3)}.
\end{align}
$\Omega_{+}$ and $\Omega_{-}$ are the prograde and retrograde angular velocities respectively. Unless otherwise stated, we always consider the prograde one in this work.  

In Fig.\ref{fig:energy}, the specific energy $\tilde{E}$ of geodesic particles on circular orbits around a short-hairy black hole is plotted as a function of orbit radius $r$. 
In the left panel, we fix the two parameters $Q=0.5, k=1.5$ and show the impact of rotation parameter $a$ on the specific energy $\tilde{E}$. As the black hole rotation parameter $a$ increases, the conserved specific energy $\tilde{E}$ of particles on circular orbits decreases. The left endpoint of each colored curve in the panel corresponds to the specific energy for particles on the innermost stable circular orbit (ISCO). In the middle panel, we fix the two parameters $a=0.5, k=1.5$ and show the impact of short-hair parameter $Q$ on the specific energy $\tilde{E}$.  In the right panel, we fix the two parameters $a=0.5, Q=0.5$ and show the impact of power parameter $k$ on the specific energy $\tilde{E}$. 
Compared with rotation parameter $a$, parameters $Q,k$ have relatively insignificant impact on the specific energy of particles on circular orbits.

In Fig.\ref{fig:angm}, we show the specific angular momenta of particles on circular orbits with different radii $r$. Similar to the specific energy, the specific angular momenta of particles on circular orbits also increases as the radius of the orbit increases for various choices of the parameters $a,Q,k$. The impact of the model parameters $a,Q,k$ on the specific angular momenta of particles are shown by colored curves in each panel. The rotation parameter has relatively larger impact on the specific angular momentum than that of the parameters $Q,k$. From the right panel, we can observe that the parameter $k$ lead to remarkable depression on the deviation of the specific angular momenta from the Kerr case.

%图片4
\begin{figure}[]
    \centering
    \subfigure{
    \begin{minipage}{.8\linewidth}
    \centering
    \includegraphics[width=\linewidth]{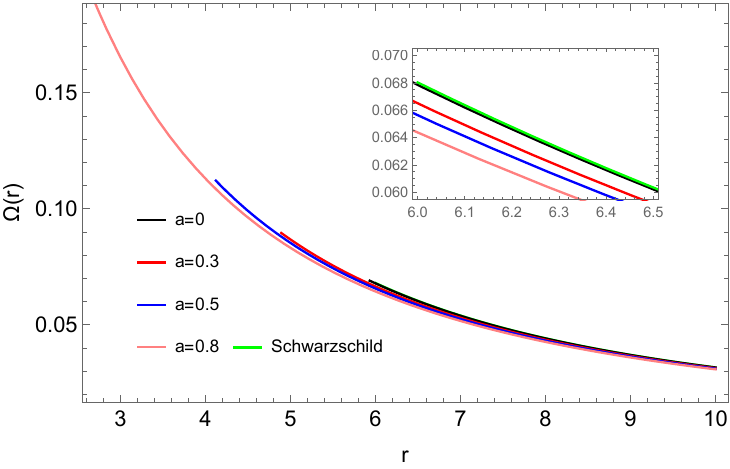}
     \end{minipage}}
     
   \subfigure{ \begin{minipage}{.8\linewidth}
    \centering
    \includegraphics[width=\linewidth]{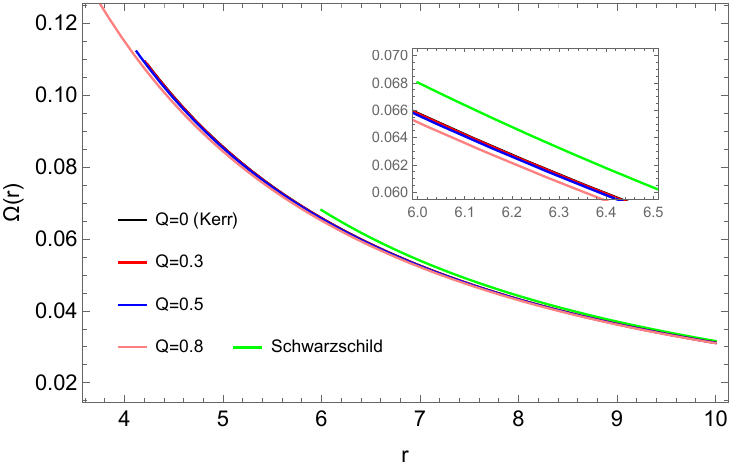}
   \end{minipage}}
   
   \subfigure{
    \begin{minipage}{.8\linewidth}
    \centering
    \includegraphics[width=\linewidth]{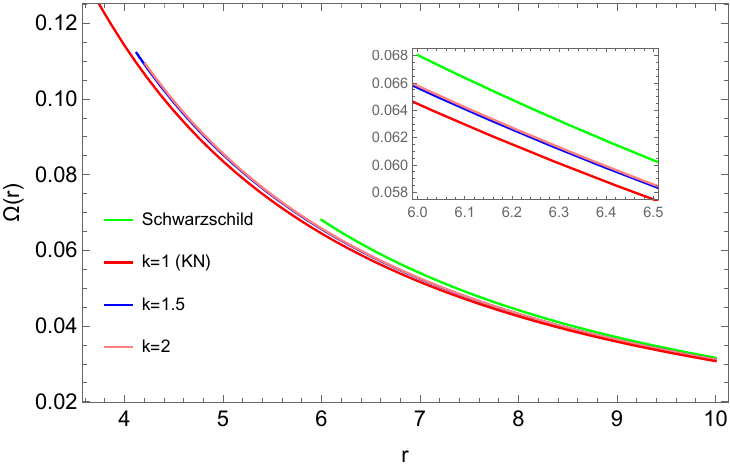}
     \end{minipage}}
       \caption{The angular velocities of particles on different circular orbits for various parameters $a,Q,k$. The parameters are chosen as: $Q=0.5, k=1.5$ for the top panel, $a=0.5, k=1.5$ for the middle panel and $a=0.5, Q=0.5$ for the bottom panel.}
    \label{fig:angv}
\end{figure}
In Fig.\ref{fig:angv}, we show the  angular velocities of particles on circular orbits with different radii. We can see that the angular velocities of particles monotonically decrease as the radius of the orbit increases for various choices of the parameters $a,Q,k$. 
And the parameters $a,Q,k$ all have insignificant impact on the angular velocities of particles.

%图片7
\begin{figure}[]
    \centering
    \subfigure{
    \begin{minipage}{.77\linewidth}
    \centering
    \includegraphics[width=\linewidth]{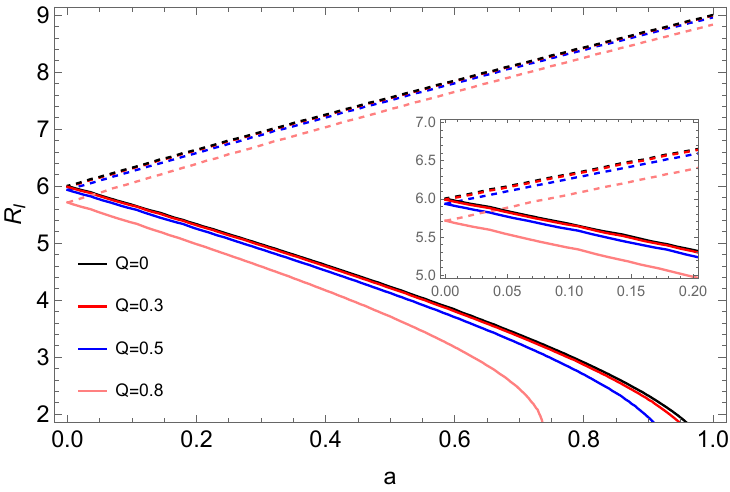}
     \end{minipage}}
     
   \subfigure{ \begin{minipage}{.77\linewidth}
    \centering
    \includegraphics[width=\linewidth]{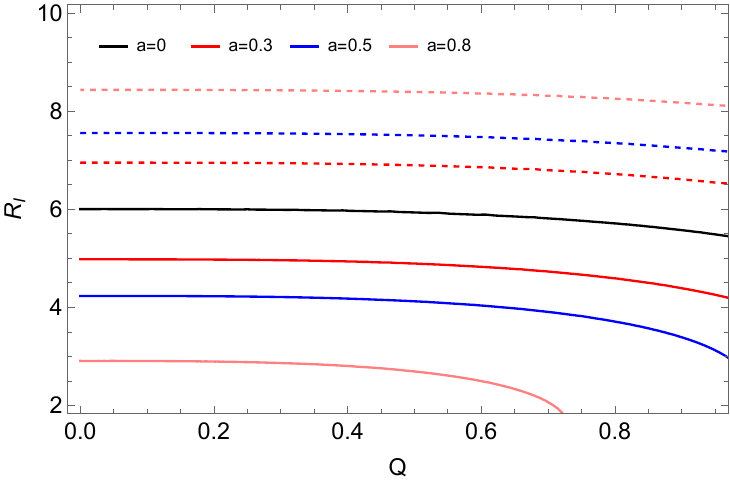}
   \end{minipage}}
   
   \subfigure{
    \begin{minipage}{.77\linewidth}
    \centering
    \includegraphics[width=\linewidth]{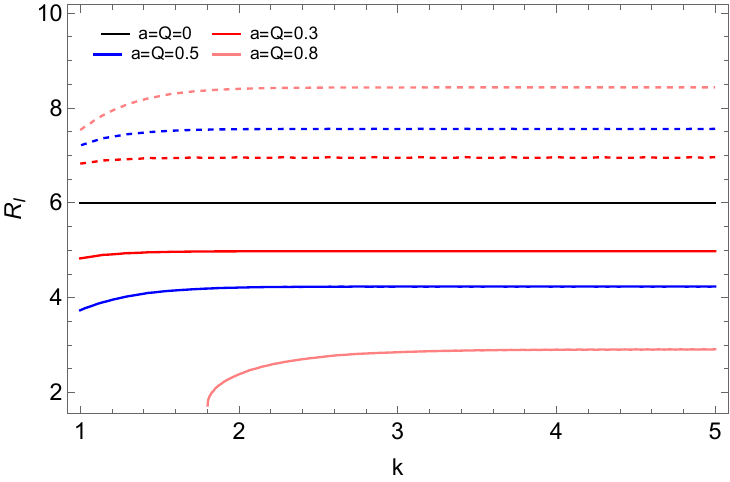}
     \end{minipage}}
       \caption{The impact of parameters $a,Q,k$ on the ISCO radii. In the top and middle panels, the power parameter $k$ is chosen to be $1.5$. The solid lines and dotted lines correspond to prograde and retrograde cases respectively. }
    \label{fig5}
\end{figure}

Not all circular orbits in the equatorial plane of short-hairy black holes are stable, and an innermost stable circular orbit (ISCO) usually exists. 
The radius of ISCO, $R_I$, is determined by equations \eqref{equation:enter-labe113}, \eqref{equation:enter-labe114} and the following additional equation
\begin{equation}
     V_{\mathrm{e f f}}^{\prime\prime}(r)=0.
\end{equation}
For given model parameters, the ISCO radii of short-hairy black holes can be solved numerically. The relations between $R_{I}$ and the parameters $a,Q,k$ are plotted in Fig.\ref{fig5}. In each panel, we plot the ISCO radii for both the prograde (solid lines) and retrograde (dotted lines) cases.  
In the top panel, we take $k=1.5$ and show how the parameter $a$ impact $R_I$. As expected, $R_I$ decreases with the increasing of $a$ for the prograde case while for retrograde case it increases. In the middle panel, $k=1.5$ and the impact of short hair parameter $Q$ on $R_I$ is shown. Since $Q$ represents a short hair, the deviation of $R_I$ from the Kerr case is visible only when $Q$ is large enough. In the bottom panel, we show the impact of the power parameter $k$ on $R_I$. Short hair condition requires $k>1$. When $k$ is larger, the hair is shorter in the radial direction. Thus, the ISCO radii are nearly constant when $k$ is large enough.

\begin{figure}[]
    \centering
    \subfigure{
    \begin{minipage}{.78\linewidth}
    \centering
    \includegraphics[width=\linewidth]{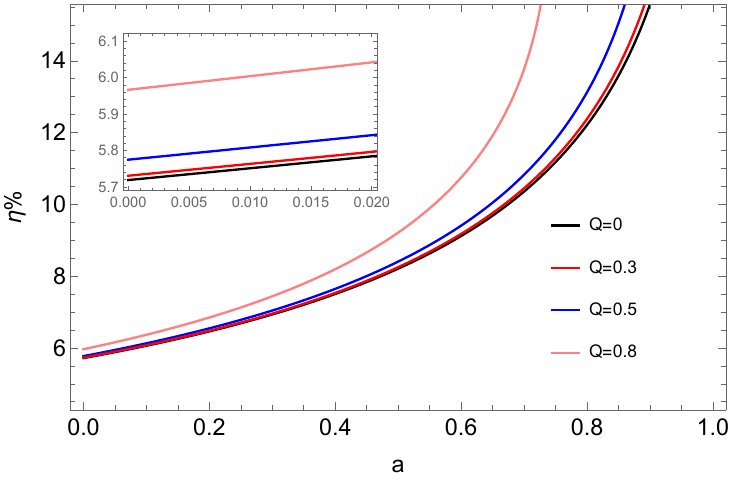}
     \end{minipage}}
     
   \subfigure{ \begin{minipage}{.78\linewidth}
    \centering
    \includegraphics[width=\linewidth]{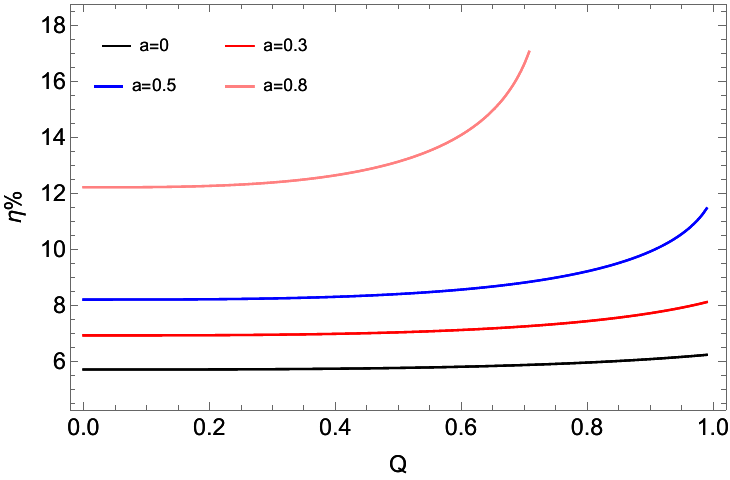}
   \end{minipage}}
   
   \subfigure{
    \begin{minipage}{.78\linewidth}
    \centering
    \includegraphics[width=\linewidth]{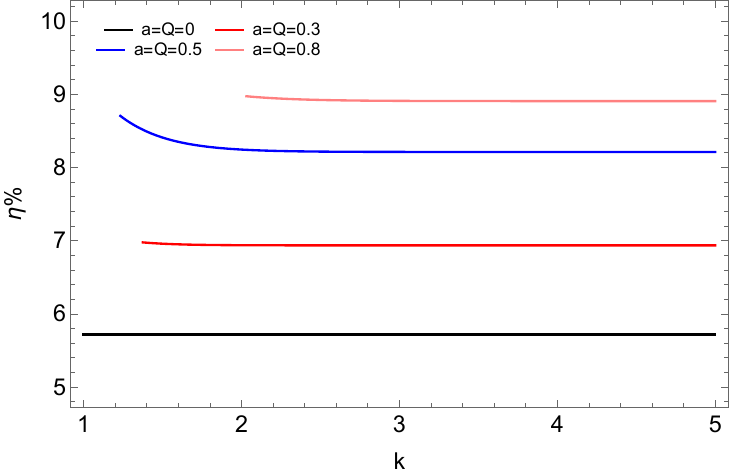}
     \end{minipage}}
       \caption{The impact of parameters $a,Q,k$ on the radiative efficiency $\eta$. In the top and middle panels, $k=1.5$.}
    \label{fig6}
\end{figure}
An important quantity related with energy of particles on ISCO is the radiative efficiency of a black hole, which characterizes the ratio between the energy radiated away
during the accretion process and the initial energy of the accretion matter \cite{Page:1974he,Collodel:2021gxu,Wu:2024sng,Kurmanov:2024hpn,Liu:2024brf,Lee:2022rtg,Asukula:2023akj,Olmo:2023lil}. 
Consider that a particle of unit mass goes from infinity to the ISCO, and assume that all lost energy is converted to radiative energy and reaches to infinity.  	 
Then, the radiative efficiency $\eta$ is defined as 
	    \begin{align}
		\eta=\frac{\tilde{E}_{\infty}-\tilde{E}_{\text{ISCO}}}{\tilde{E}_{\infty}}\approx 1-\tilde{E}_{\text{ISCO}},
		\end{align}
where we use  $\tilde{E}_{\infty}\approx1$. The impact of the parameters $a,Q,k$ on the radiative efficiencies of short-hairy black holes are shown in Fig.\ref{fig6}. 
From the top panel, we can see that the rotation parameter $a$ improve the radiative efficiency significantly. In the middle panel, we can see that although the radiative efficiencies increase as the short hair parameter $Q$ increases, the improvement of the efficiency is visible only when $Q$ is large enough. 
Especially, for highly rotating hairy black holes, the radiative efficiency has a remarkable increase for large $Q$.
From the bottom panel, we see that the parameter $k$ generally have less impact on the radiative efficiency.

\section{Radiative Properties of thin disks around rotating short-hairy black holes}\label{sec.4}
\begin{figure*}[]
    \centering
    \subfigure{
    \begin{minipage}{.31\linewidth}
    \centering
    \includegraphics[width=\linewidth]{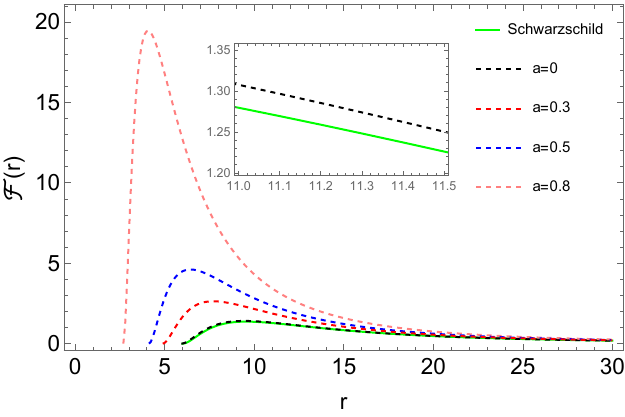}
     \end{minipage}}
     \subfigure{ \begin{minipage}{.31\linewidth}
    \centering
    \includegraphics[width=\linewidth]{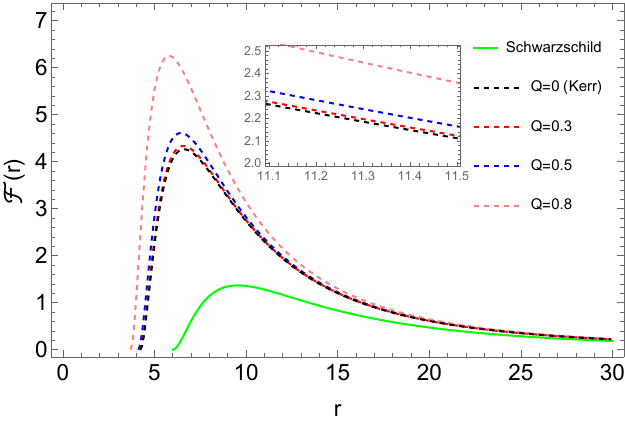}
   \end{minipage}}
   \subfigure{
    \begin{minipage}{.31\linewidth}
    \centering
    \includegraphics[width=\linewidth]{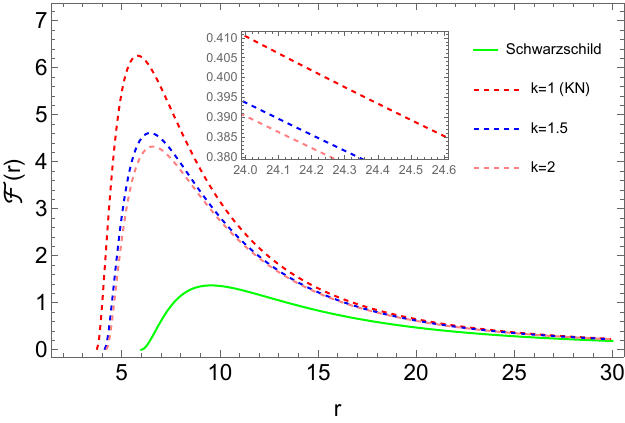}
     \end{minipage}} 
   \caption{Radiative flux densities (multiplied by $10^5$) as functions of orbit radius $r$. In the left panel, $Q=0.5,k=1.5$. In the middle panel, $a=0.5,k=1.5$. In the right panel, $a=0.5,Q=0.5$.}
    \label{fig7}
\end{figure*}

In this section, we focus on the radiative properties of a thin accretion disk around the short-hairy black hole and investigate how the model parameters impact on the properties.
In the spherical coordinates, the radiative flux density (the energy per unit area per unit time) emitted by the accretion disk is defined as \cite{Page:1974he,Collodel:2021gxu,Wu:2024sng},
	\begin{equation}\label{flux}
		 \mathcal{F}(r)=-\frac{\dot{M}}{4\pi\sqrt{-g/g_{\theta\theta}}}\frac{\Omega_{,r}}{(\tilde{E}-\Omega \tilde{L})^{2}}\int_{R_I}^{r}(\tilde{E}-\Omega \tilde{L})\tilde{L}_{,r}dr,
	\end{equation}
where $\dot{M}$ is the mass accretion rate which is assumed a constant here, and $g$ is the determinant of the metric, which is $ g=g_{r r}g_{\theta\theta}(g_{t t} g_{\phi \phi}-g_{t \phi}^2)$ in our case. 

The radiative flux densities per unit accretion rate are plotted in Fig.\ref{fig7} with different specific values of the parameters $a,Q,k$. The impact of the parameters on the radiative flux densities are shown in different panels. We can see that there always exists an maximum for the radiative flux density of a short-hairy black hole. In the left panel, we can see that the radiative flux densities increase as the rotation parameter $a$ increases. The location of the maximum of the radiative flux density decreases as $a$ increases. The rotation parameter $a$ can improve the maximum remarkably. When $Q=0.5,k=1.5$, the maximum for $a=0.8$ is about four times the maximum for $a=0.5$. In the middle panel, we can see that the radiative flux density has visible deviation from the Kerr case only when the short hair parameter $Q$ is large enough. When $a=0.5,k=1.5$, the maximum of the radiative flux density increases about $6\%$ for $Q=0.5$ and about $45\%$ for $Q=0.8$, compared to the Kerr case. In the right panel, the power parameter $k$ can depress the deviation from the Kerr case significantly.

Assume the accretion disk is in local thermodynamic equilibrium and treat its radiation as a black body radiation, then the radiative flux density $\mathcal{F}$ and the local effective temperature of the disk $T_{\mathrm{eff}}$ satisfy the Stefan-Boltzmann law \cite{Nampalliwar:2018tup,Jiang:2024njc}, i.e.,
\begin{equation}
   \mathcal{F}= \sigma T^4_{\mathrm{eff}},
\end{equation}
where $\sigma=\frac{2\pi^{5}k^{4}}{15c^{2}h^{3}}$ is the Stefan–Boltzmann constant.
When the radiation emitted by the disk reaches an observer at infinity, it is affected by gravitational and Doppler redshifts. 
Then for the observer, the disk appears to have a different temperature $T_{\infty}$.
The redshift factor $z$ in our model is \cite{Bhattacharyya:2000kt}
\begin{equation}
    1+z=\frac{1+\Omega r\sin\phi\sin\gamma}{\sqrt{-g_{tt}-2g_{t\phi}\Omega-g_{\phi\phi}\Omega^{2}}},
\end{equation}
where $\gamma$ is the inclination angle of the distant observer to the axis of the accretion disk. Thus, neglecting the light bending, the temperature detected by a distant observer is \cite{Karimov:2018whx,Joshi:2013dva,Jiang:2024njc}
\begin{equation}
  T_{\infty}=\frac{T_{\mathrm{eff}}}{1+z}.
\end{equation}

\begin{figure}[]
    \centering
    \subfigure{
    \begin{minipage}{.7\linewidth}
    \centering
    \includegraphics[width=\linewidth]{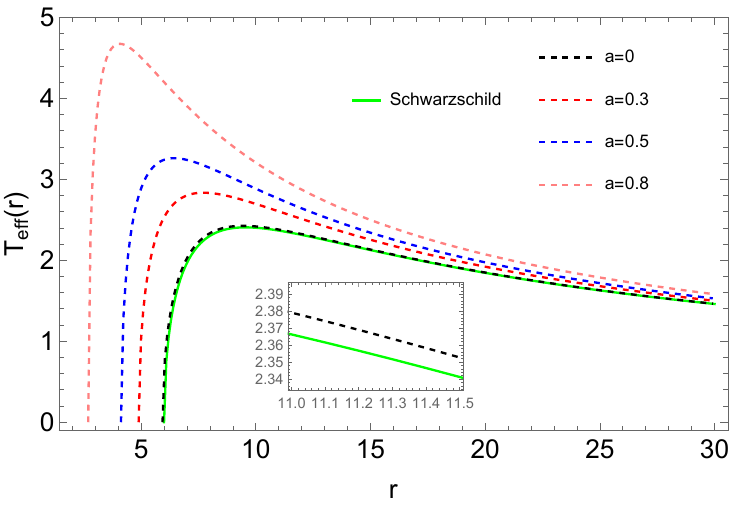}
     \end{minipage}}   
       
   \subfigure{ \begin{minipage}{.7\linewidth}
    \centering
    \includegraphics[width=\linewidth]{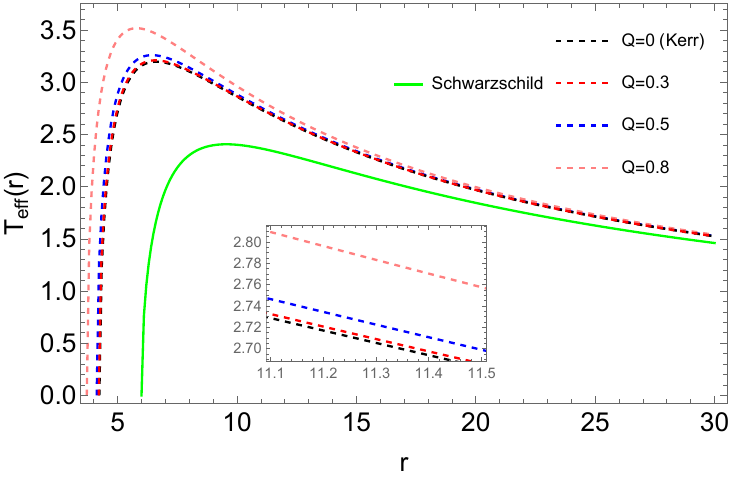}
   \end{minipage}}
   
   \subfigure{
    \begin{minipage}{.7\linewidth}
    \centering
    \includegraphics[width=\linewidth]{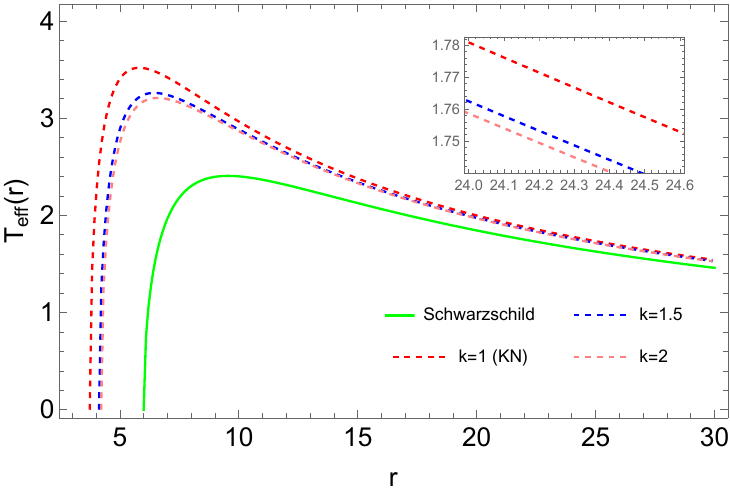}
     \end{minipage}}
     
       \caption{Local effective temperatures (multiplied by $10^2$) as functions of orbit radius $r$. In the top panel, $Q=0.5,k=1.5$. In the middle panel, $a=0.5,k=1.5$. In the bottom panel, $a=0.5,Q=0.5$. }
    \label{figtemp}
\end{figure}
The local effective temperatures are plotted in Fig.\ref{figtemp} with different specific values of the parameters $a,Q,k$. The impact of parameters on the effective temperatures are shown in different panels. For a short-hairy black hole, the local temperature of its disk always has a maximum. In the top panel, we can see that the effective temperature increases as the rotation parameter $a$ increases. 
The location of the maximum of the effective temperature decreases as $a$ increases and the value of the maximum increase remarkably. For a short-hairy black hole with $Q=0.5,k=1.5$, the maximum for $a=0.8$ is about 1.6 times the maximum for $a=0.3$. In the middle panel, we can see that the effective temperature also increases as the short hair parameter $Q$ increases. For a short-hairy black hole with $a=0.5,k=1.5$, the maximum for $Q=0.8$ is about 1.1 times the maximum for $Q=0.3$. In the bottom panel, we can see the power parameter $k$ depresses the impact of the short hair remarkably.   

\begin{figure}[]
    \centering
    \includegraphics[width=.75\linewidth]{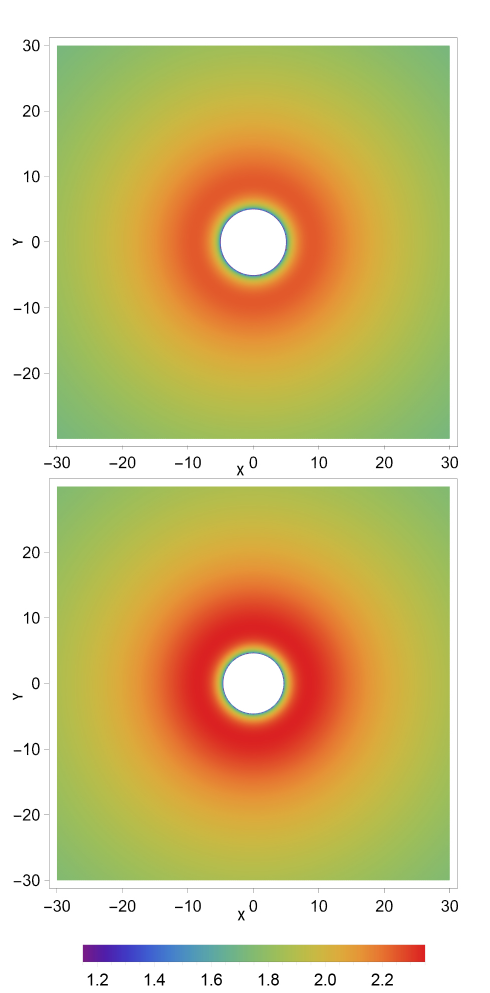}
       \caption{The temperature (multiplied by $10^2$) of a short-hairy black hole detected by a distant observer (bottom) and that of a Kerr black hole (top). }
    \label{fig:density}
\end{figure}

For a distant observer at a position with an inclination angle $\gamma=0$, taking into account the redshift factor, the observed temperatures of a short-hairy black hole with $a=0.3,Q=0.8,k=1.5$ (the bottom one) and a Kerr black hole with $a=0.3$ (the top one) are plotted in Fig.\ref{fig:density}. In the accretion disk plane, $X=r\cos\phi,Y=r\sin\phi$.  
We can observe that the short hair leads to the increase of temperature of the accretion disk near the black hole, compared to the Kerr case. 

%\begin{figure}[]
%    \centering
%    \subfigure{
%    \begin{minipage}{.31\linewidth}
%    \centering
%    \includegraphics[width=\linewidth]{densityplotKerr.pdf}
%     \end{minipage}}     
%   \subfigure{ \begin{minipage}{.31\linewidth}
%    \centering
%    \includegraphics[width=\linewidth]{densityplotKerr.pdf}
%   \end{minipage}}   
%   \caption{The relationship between r and  differential of the luminosity,black-body spectrum}
%    \label{fig:density}
%\end{figure}

It is worth noting that the radiative flux density is not directly observable, since it is a quantity measured in the rest frame of the accretion disk.
A more practical quantity is the luminosity $\mathcal{L}_{\infty}$ that reaches an observer at infinity \cite{Joshi:2013dva}.
The differential of the luminosity $\mathcal{L}_{\infty}$ with respect to $r$ is defined as \cite{Joshi:2013dva,Boshkayev:2023fft}
 \begin{equation}
    \frac{\mathrm{d}\mathcal{L}_{\infty}}{\mathrm{d}\ln r}=4 \pi r \sqrt{-g} \tilde{E} \mathcal{F}(r).
\end{equation}

%$ \frac{\mathrm{d}\mathcal{L}_{\infty}}{\mathrm{d}\ln r}\sim r^\alpha$, plot $\log(\frac{\mathrm{d}\mathcal{L}_{\infty}}{\mathrm{d}\ln r})\sim \log(r)$. 

From the differential of luminosity $\mathcal{L}_{\infty}$, we can obtain the spectral luminosity distribution $\nu\mathcal{L}_{\nu,\infty}$ observed at infinity, which is calculated as \cite{Joshi:2013dva,Boshkayev:2023fft,Jiang:2024njc}
%\begin{align}
%\nu\mathcal{L}_{\nu,\infty}&=\frac{15}{\pi^{4}}\int^{\infty}_{r_I} \left(\frac{\mathrm{d}\mathcal{L}_{\infty}}{\mathrm{d}\ln r}\right)\nonumber\\
%&\times\frac{(1+z)^4(h\nu/(k T_{\mathrm{eff}}))^4}{\exp[(1+z)(h\nu/(k T_{\mathrm{eff}}))]-1}\mathrm{d}\ln r.
%\end{align}

\begin{equation}
\nu\mathcal{L}_{\nu,\infty}=\frac{8\pi h\cos \gamma}{c^{2}}\int_{r_I}^{r_{\mathrm{out}}}\int_{0}^{2\pi}\frac{\nu\nu_{e}^{3}r\mathrm{d}r\mathrm{d}\phi}{\exp(\frac{h\nu_{e}}{kT_{\mathrm{eff}}})-1},
\end{equation}
where $\nu_e$ is the radiative frequency in the local rest frame of the disk, which is related to the observed frequency $\nu$ at infinity by $\nu_e=\nu(1+z)$.
$r_{\mathrm{out}}$ is the outer edge of the accretion disk, which is chosen to be $100M$ here.

\begin{figure*}[tbph]
    \centering
    \subfigure{
    \begin{minipage}{.31\linewidth}
    \centering
    \includegraphics[width=\linewidth]{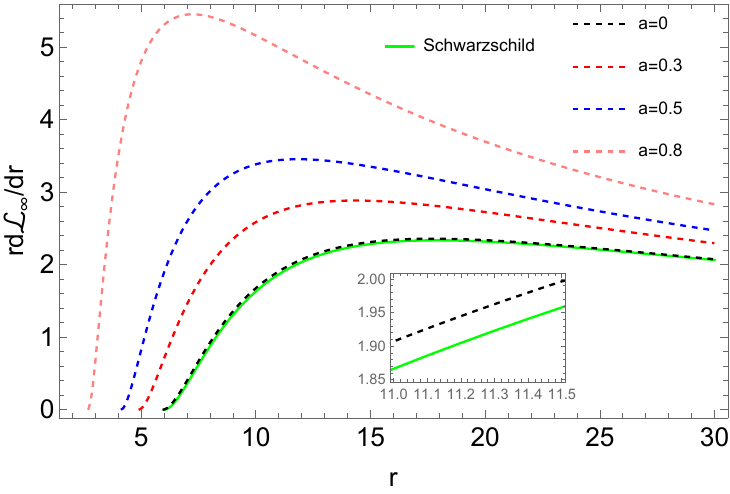}
     \end{minipage}}     
   \subfigure{ \begin{minipage}{.31\linewidth}
    \centering
    \includegraphics[width=\linewidth]{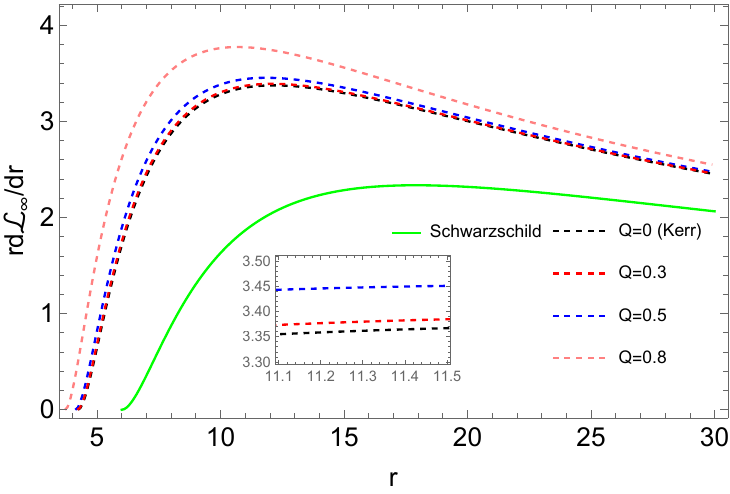}
   \end{minipage}}      
   \subfigure{
    \begin{minipage}{.31\linewidth}
    \centering
    \includegraphics[width=\linewidth]{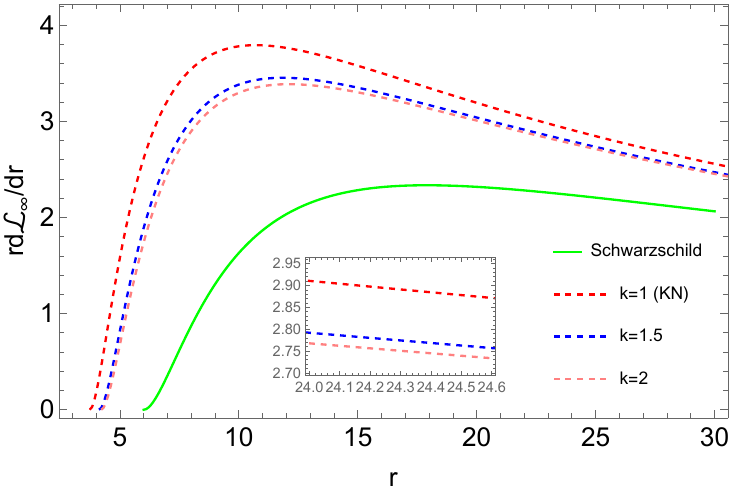}
     \end{minipage}}
    \caption{Differential of luminosity per unit mass accretion (multiplied by $10^2$) and impact of parameters $a,Q,k$. In the left panel, $Q=0.5,k=1.5$. In the middle panel, $a=0.5,k=1.5$. In the right panel, $a=0.5,Q=0.5$.}
    \label{fig:luminosity}
\end{figure*}

The differential of luminosity for short-hairy black holes with various parameters are plotted in Fig.\ref{fig:luminosity}. The influences of parameters on the differential of luminosity are shown in different panels.
 For a short-hairy black hole, the differential of luminosity always has a maximum. From the left panel, we can observe that the maximum of the differential of luminosity increases as the rotation parameter $a$ increases while the location of the maximum decreases. For a short-hairy black hole with $Q=0.5,k=1.5$, the maximum for $a=0.8$ is about 2 times the maximum for $a=0.3$. From the middle panel, we can see that the differential of luminosity also increases as the short hair parameter $Q$ increases.
For a short-hairy black hole with $a=0.5,k=1.5$, the maximum for $Q=0.8$ is about 1.1 times the maximum for $Q=0.3$.
From the right panel, we can see that the differential of luminosity decreases as the power parameter $k$ increases.

\begin{figure*}[tbph]
    \centering
    \subfigure{
    \begin{minipage}{.31\linewidth}
    \centering
    \includegraphics[width=\linewidth]{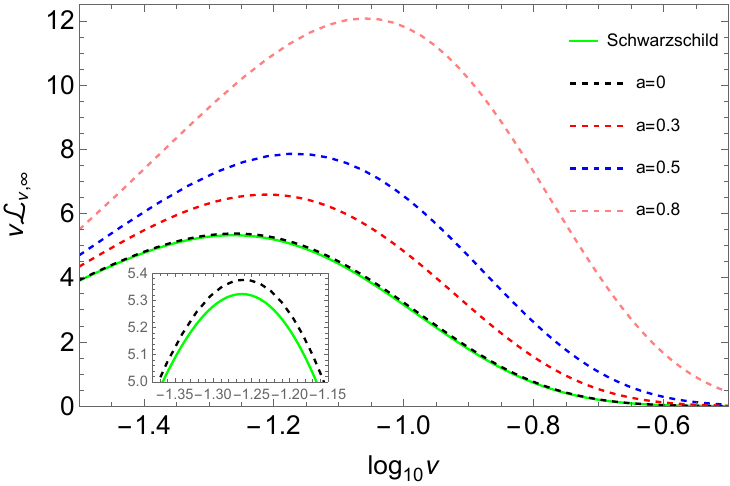}
     \end{minipage}} 
      \subfigure{ \begin{minipage}{.31\linewidth}
    \centering
    \includegraphics[width=\linewidth]{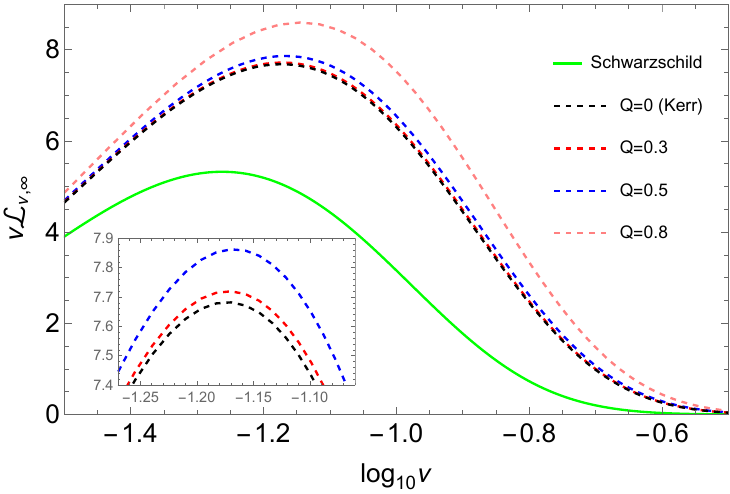}
   \end{minipage}}
   \subfigure{
    \begin{minipage}{.31\linewidth}
    \centering
    \includegraphics[width=\linewidth]{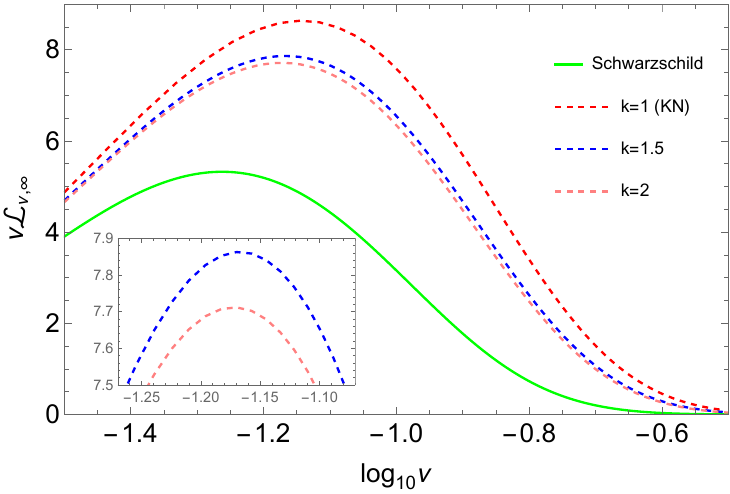}
     \end{minipage}}
       \caption{Spectral luminosity distribution (multiplied by $10^2$) and impact of parameters $a,Q,k$. In the left panel, $Q=0.5,k=1.5$. In the middle panel, $a=0.5,k=1.5$. In the right panel, $a=0.5,Q=0.5$.}
    \label{fig:spectral}
\end{figure*}

The spectral luminosity distribution for short-hairy black holes with various parameters are plotted in Fig.\ref{fig:spectral}.  For a short-hairy black hole, the spectral luminosity distribution always has a maximum.
In the left panel, we observe that the rotation parameter $a$ increases the spectral luminosity distribution remarkably. For a short-hairy black hole with $Q=0.5,k=1.5$, the maximum for $a=0.8$ is about 1.8 times the maximum for $a=0.3$. The frequency corresponding to the maximum increases as the rotation parameter $a$ increases. From the middle panel, we can see that the spectral luminosity distribution also increases as the short hair parameter $Q$ increases.
For a short-hairy black hole with $a=0.5,k=1.5$, the maximum for $Q=0.8$ is about 1.1 times the maximum for $Q=0.3$. From the right panel, we can see the power parameter $k$ depresses the impact of the short hair on the spectral luminosity distribution remarkably.

%图片11

\section{Summary}
The focus of this work is the radiative properties of thin accretion disks around the rotating short-hairy black holes and the impact of the rotation parameter $a$, short-hair parameter $Q$ and power parameter $k$ on these properties. After introducing the short-hairy black holes, we study the geodesic motion of massive particles in the equatorial planes. The radius of the innermost stable circular orbit is derived. The conserved energy and angular momenta are also derived for particles on different circular orbits. 
The impact of the model parameters on these quantities are numerically plotted in figures. The radiative efficiencies $\eta$ of the short-hairy black holes with various parameters are also calculated. Compared to the Kerr case, it is found that the short hair $Q$ enhances the radiative efficiencies, especially when the short hair is large enough. 

The various aspects of the radiative properties of thin accretion disks around the rotating short-hairy black holes are then considered. The radiative flux densities and effective temperatures of the disks with various parameters are plotted in figures. It is found that the short hair can enhance the radiation and increase the temperature of the disk. Besides the two quantities defined in local rest frame of the disk, we also show the temperature detected by a distant observer by taking into account the redshift effects and compare it with the corresponding Kerr case. The differential of luminosity and spectral luminosity distribution are also plotted for various specific values of the model parameters.    
Compared to the Kerr case, these quantities are also enhanced by the short hair of the black hole. 

For the radiative properties of thin accretion disks around the rotating short-hairy black holes, compared to the parameters $Q, k$, the rotation parameter $a$ has more remarkable impact on them. As the power parameter $k$ increases, the impact of the short hair on these properties is depressed significantly. For moderate power parameter $k\sim 1.5$, when the short hair parameter $Q$ is large enough($\sim 0.8$), it leads to remarkable impact on these properties. 

Although the results in this work focuses on theoretical aspect, we hope they may provide preliminaries for distinguishing Kerr and short-hairy black holes in future observations \cite{Bambi:2015kza,Bambi:2020cyv}.

\section{appendix}
The specific energy $\tilde{E}$ and specific angular momentum $\tilde{L}$ are obtained by solving the equations: $V_{\mathrm{eff}}(r)=0,~ V_{\mathrm{eff}}^{\prime}(r)=0$. The explicit expressions for them are as follows:
\begin{align}
\tilde{L}=&(k ra^3 Q^{2 k}+B a^2r^{2k+2}+2Ba^2r^{2k+1}-a^3 r^{2 k}\nonumber\\
&-Ba^2r^2Q^{2 k}+akr^3Q^{2 k}+ar^3Q^{2 k}-3ar^{2 k+2}\nonumber\\
&+Br^{2 k+4} ) (-2kBa^3 r^{2 k+3} Q^{2 k}+2a^3 r^{4 k+2} \nonumber\\
& +a^2 (-k^2 r^4 Q^{4 k}+Q^{2 k} r^{2 k+3}+ 3 kr^{2 k+4} Q^{2 k}\nonumber\\
&+4k  Q^{2 k} r^{2 k+3}-k r^4 Q^{4 k}-3r^{4 k+3}+3 r^{4 k+2})\nonumber\\
&+r^{2 k+5} (kr Q^{2 k}+r Q^{2 k}+r^{2 k+1}-3 r^{2 k})\nonumber\\
&+2 aB r^{2 k+4} (3 r^{2 k}-kr Q^{2 k}+ r Q^{2 k}) )^{-1/2},
 \end{align}

\begin{equation}
\begin{aligned}
\tilde{E}=&(kra^2 Q^{2 k}-a^2r^{2 k}+2aBr^{2 k+1}+r^{2 k+3}\\&-aBr^2 Q^{2 k}+r^3 Q^{2 k}-2 r^{2 k+2})(2 Ba^3 r^{4 k+2}\\&-2 kBa^3 r^{2 k+3} Q^{2 k}+a^2 r^2 ( 3kQ^{2 k} r^{2 k+2}\\&+4kQ^{2 k} r^{2 k+1}+Q^{2 k} r^{2 k+1}-k^2r^2 Q^{4 k}\\&+k r^2 Q^{4 k}-3 r^{4 k+1}+ r^{4 k})+2 aB r^{2 k+4}\\& (3 r^{2 k}-kr Q^{2 k}- r Q^{2 k})+kQ^{2 k} r^{2 k+6}\\&+Q^{2 k} r^{2 k+6}+r^{4 k+6}-3r^{4 k+5})^{-1/2},
\end{aligned}
\end{equation}
where 
\begin{equation}
   B=\sqrt{\frac{1}{r}-\frac{kQ^{2k}}{r^{2k}}}.
\end{equation}

\section*{Acknowledgments}
This work is partially supported by Guangdong Major Project of Basic and Applied Basic Research (No.2020B0301030008).

\bibliography{refshthair5.bib}

%apsrev4-2.bst 2019-01-14 (MD) hand-edited version of apsrev4-1.bst
%Control: key (0)
%Control: author (8) initials jnrlst
%Control: editor formatted (1) identically to author
%Control: production of article title (0) allowed
%Control: page (0) single
%Control: year (1) truncated
%Control: production of eprint (0) enabled
\begin{thebibliography}{103}%
\makeatletter
\providecommand \@ifxundefined [1]{%
 \@ifx{#1\undefined}
}%
\providecommand \@ifnum [1]{%
 \ifnum #1\expandafter \@firstoftwo
 \else \expandafter \@secondoftwo
 \fi
}%
\providecommand \@ifx [1]{%
 \ifx #1\expandafter \@firstoftwo
 \else \expandafter \@secondoftwo
 \fi
}%
\providecommand \natexlab [1]{#1}%
\providecommand \enquote  [1]{``#1''}%
\providecommand \bibnamefont  [1]{#1}%
\providecommand \bibfnamefont [1]{#1}%
\providecommand \citenamefont [1]{#1}%
\providecommand \href@noop [0]{\@secondoftwo}%
\providecommand \href [0]{\begingroup \@sanitize@url \@href}%
\providecommand \@href[1]{\@@startlink{#1}\@@href}%
\providecommand \@@href[1]{\endgroup#1\@@endlink}%
\providecommand \@sanitize@url [0]{\catcode `\\12\catcode `\$12\catcode
  `\&12\catcode `\#12\catcode `\^12\catcode `\_12\catcode `\%12\relax}%
\providecommand \@@startlink[1]{}%
\providecommand \@@endlink[0]{}%
\providecommand \url  [0]{\begingroup\@sanitize@url \@url }%
\providecommand \@url [1]{\endgroup\@href {#1}{\urlprefix }}%
\providecommand \urlprefix  [0]{URL }%
\providecommand \Eprint [0]{\href }%
\providecommand \doibase [0]{https://doi.org/}%
\providecommand \selectlanguage [0]{\@gobble}%
\providecommand \bibinfo  [0]{\@secondoftwo}%
\providecommand \bibfield  [0]{\@secondoftwo}%
\providecommand \translation [1]{[#1]}%
\providecommand \BibitemOpen [0]{}%
\providecommand \bibitemStop [0]{}%
\providecommand \bibitemNoStop [0]{.\EOS\space}%
\providecommand \EOS [0]{\spacefactor3000\relax}%
\providecommand \BibitemShut  [1]{\csname bibitem#1\endcsname}%
\let\auto@bib@innerbib\@empty
%</preamble>
\bibitem [{\citenamefont {Abbott}\ \emph {et~al.}(2016)\citenamefont {Abbott}
  \emph {et~al.}}]{LIGOScientific:2016aoc}%
  \BibitemOpen
  \bibfield  {author} {\bibinfo {author} {\bibfnamefont {B.~P.}\ \bibnamefont
  {Abbott}} \emph {et~al.} (\bibinfo {collaboration} {LIGO Scientific,
  Virgo}),\ }\bibfield  {title} {\bibinfo {title} {{Observation of
  Gravitational Waves from a Binary Black Hole Merger}},\ }\href
  {https://doi.org/10.1103/PhysRevLett.116.061102} {\bibfield  {journal}
  {\bibinfo  {journal} {Phys. Rev. Lett.}\ }\textbf {\bibinfo {volume} {116}},\
  \bibinfo {pages} {061102} (\bibinfo {year} {2016})},\ \Eprint
  {https://arxiv.org/abs/1602.03837} {arXiv:1602.03837 [gr-qc]} \BibitemShut
  {NoStop}%
\bibitem [{\citenamefont {Akiyama}\ \emph {et~al.}(2019)\citenamefont {Akiyama}
  \emph {et~al.}}]{M87-EHT-1}%
  \BibitemOpen
  \bibfield  {author} {\bibinfo {author} {\bibfnamefont {K.}~\bibnamefont
  {Akiyama}} \emph {et~al.} (\bibinfo {collaboration} {Event Horizon
  Telescope}),\ }\bibfield  {title} {\bibinfo {title} {{First M87 Event Horizon
  Telescope Results. I. The Shadow of the Supermassive Black Hole}},\ }\href
  {https://doi.org/10.3847/2041-8213/ab0ec7} {\bibfield  {journal} {\bibinfo
  {journal} {Astrophys. J. Lett.}\ }\textbf {\bibinfo {volume} {875}},\
  \bibinfo {pages} {L1} (\bibinfo {year} {2019})},\ \Eprint
  {https://arxiv.org/abs/1906.11238} {arXiv:1906.11238 [astro-ph.GA]}
  \BibitemShut {NoStop}%
\bibitem [{\citenamefont {Akiyama}\ \emph {et~al.}(2022)\citenamefont {Akiyama}
  \emph {et~al.}}]{SgrA-EHT-1}%
  \BibitemOpen
  \bibfield  {author} {\bibinfo {author} {\bibfnamefont {K.}~\bibnamefont
  {Akiyama}} \emph {et~al.} (\bibinfo {collaboration} {Event Horizon
  Telescope}),\ }\bibfield  {title} {\bibinfo {title} {{First Sagittarius A*
  Event Horizon Telescope Results. I. The Shadow of the Supermassive Black Hole
  in the Center of the Milky Way}},\ }\href
  {https://doi.org/10.3847/2041-8213/ac6674} {\bibfield  {journal} {\bibinfo
  {journal} {Astrophys. J. Lett.}\ }\textbf {\bibinfo {volume} {930}},\
  \bibinfo {pages} {L12} (\bibinfo {year} {2022})},\ \Eprint
  {https://arxiv.org/abs/2311.08680} {arXiv:2311.08680 [astro-ph.HE]}
  \BibitemShut {NoStop}%
\bibitem [{\citenamefont {Kerr}(1963)}]{Kerr:1963ud}%
  \BibitemOpen
  \bibfield  {author} {\bibinfo {author} {\bibfnamefont {R.~P.}\ \bibnamefont
  {Kerr}},\ }\bibfield  {title} {\bibinfo {title} {{Gravitational field of a
  spinning mass as an example of algebraically special metrics}},\ }\href
  {https://doi.org/10.1103/PhysRevLett.11.237} {\bibfield  {journal} {\bibinfo
  {journal} {Phys. Rev. Lett.}\ }\textbf {\bibinfo {volume} {11}},\ \bibinfo
  {pages} {237} (\bibinfo {year} {1963})}\BibitemShut {NoStop}%
\bibitem [{\citenamefont {Ruffini}\ and\ \citenamefont
  {Wheeler}(1971)}]{Ruffini:1971bza}%
  \BibitemOpen
  \bibfield  {author} {\bibinfo {author} {\bibfnamefont {R.}~\bibnamefont
  {Ruffini}}\ and\ \bibinfo {author} {\bibfnamefont {J.~A.}\ \bibnamefont
  {Wheeler}},\ }\bibfield  {title} {\bibinfo {title} {{Introducing the black
  hole}},\ }\href {https://doi.org/10.1063/1.3022513} {\bibfield  {journal}
  {\bibinfo  {journal} {Phys. Today}\ }\textbf {\bibinfo {volume} {24}},\
  \bibinfo {pages} {30} (\bibinfo {year} {1971})}\BibitemShut {NoStop}%
\bibitem [{\citenamefont {Robinson}(1975)}]{Robinson:1975bv}%
  \BibitemOpen
  \bibfield  {author} {\bibinfo {author} {\bibfnamefont {D.~C.}\ \bibnamefont
  {Robinson}},\ }\bibfield  {title} {\bibinfo {title} {{Uniqueness of the Kerr
  black hole}},\ }\href {https://doi.org/10.1103/PhysRevLett.34.905} {\bibfield
   {journal} {\bibinfo  {journal} {Phys. Rev. Lett.}\ }\textbf {\bibinfo
  {volume} {34}},\ \bibinfo {pages} {905} (\bibinfo {year} {1975})}\BibitemShut
  {NoStop}%
\bibitem [{\citenamefont {Hawking}(1972)}]{Hawking:1971vc}%
  \BibitemOpen
  \bibfield  {author} {\bibinfo {author} {\bibfnamefont {S.~W.}\ \bibnamefont
  {Hawking}},\ }\bibfield  {title} {\bibinfo {title} {{Black holes in general
  relativity}},\ }\href {https://doi.org/10.1007/BF01877517} {\bibfield
  {journal} {\bibinfo  {journal} {Commun. Math. Phys.}\ }\textbf {\bibinfo
  {volume} {25}},\ \bibinfo {pages} {152} (\bibinfo {year} {1972})}\BibitemShut
  {NoStop}%
\bibitem [{\citenamefont {Carter}(1971)}]{Carter:1971zc}%
  \BibitemOpen
  \bibfield  {author} {\bibinfo {author} {\bibfnamefont {B.}~\bibnamefont
  {Carter}},\ }\bibfield  {title} {\bibinfo {title} {{Axisymmetric Black Hole
  Has Only Two Degrees of Freedom}},\ }\href
  {https://doi.org/10.1103/PhysRevLett.26.331} {\bibfield  {journal} {\bibinfo
  {journal} {Phys. Rev. Lett.}\ }\textbf {\bibinfo {volume} {26}},\ \bibinfo
  {pages} {331} (\bibinfo {year} {1971})}\BibitemShut {NoStop}%
\bibitem [{\citenamefont {Israel}(1967)}]{Israel:1967wq}%
  \BibitemOpen
  \bibfield  {author} {\bibinfo {author} {\bibfnamefont {W.}~\bibnamefont
  {Israel}},\ }\bibfield  {title} {\bibinfo {title} {{Event horizons in static
  vacuum space-times}},\ }\href {https://doi.org/10.1103/PhysRev.164.1776}
  {\bibfield  {journal} {\bibinfo  {journal} {Phys. Rev.}\ }\textbf {\bibinfo
  {volume} {164}},\ \bibinfo {pages} {1776} (\bibinfo {year}
  {1967})}\BibitemShut {NoStop}%
\bibitem [{\citenamefont {Mazur}(1982)}]{Mazur:1982db}%
  \BibitemOpen
  \bibfield  {author} {\bibinfo {author} {\bibfnamefont {P.~O.}\ \bibnamefont
  {Mazur}},\ }\bibfield  {title} {\bibinfo {title} {{Proof of uniqueness of the
  kerr-newman black hole solution}},\ }\href
  {https://doi.org/10.1088/0305-4470/15/10/021} {\bibfield  {journal} {\bibinfo
   {journal} {J. Phys. A}\ }\textbf {\bibinfo {volume} {15}},\ \bibinfo {pages}
  {3173} (\bibinfo {year} {1982})}\BibitemShut {NoStop}%
\bibitem [{\citenamefont {Gurlebeck}(2015)}]{Gurlebeck:2015xpa}%
  \BibitemOpen
  \bibfield  {author} {\bibinfo {author} {\bibfnamefont {N.}~\bibnamefont
  {Gurlebeck}},\ }\bibfield  {title} {\bibinfo {title} {{No-hair theorem for
  Black Holes in Astrophysical Environments}},\ }\href
  {https://doi.org/10.1103/PhysRevLett.114.151102} {\bibfield  {journal}
  {\bibinfo  {journal} {Phys. Rev. Lett.}\ }\textbf {\bibinfo {volume} {114}},\
  \bibinfo {pages} {151102} (\bibinfo {year} {2015})},\ \Eprint
  {https://arxiv.org/abs/1503.03240} {arXiv:1503.03240 [gr-qc]} \BibitemShut
  {NoStop}%
\bibitem [{\citenamefont {Bekenstein}(1995)}]{Bekenstein1995}%
  \BibitemOpen
  \bibfield  {author} {\bibinfo {author} {\bibfnamefont {J.~D.}\ \bibnamefont
  {Bekenstein}},\ }\bibfield  {title} {\bibinfo {title} {{Novel
  \textquoteleft{}\textquoteleft{}no-scalar-hair\textquoteright{}\textquoteright{}
  theorem for black holes}},\ }\href
  {https://doi.org/10.1103/PhysRevD.51.R6608} {\bibfield  {journal} {\bibinfo
  {journal} {Phys. Rev. D}\ }\textbf {\bibinfo {volume} {51}},\ \bibinfo
  {pages} {R6608} (\bibinfo {year} {1995})}\BibitemShut {NoStop}%
\bibitem [{\citenamefont {Ghosh}\ \emph {et~al.}(2023)\citenamefont {Ghosh},
  \citenamefont {Sk},\ and\ \citenamefont {Sarkar}}]{Ghosh:2023kge}%
  \BibitemOpen
  \bibfield  {author} {\bibinfo {author} {\bibfnamefont {R.}~\bibnamefont
  {Ghosh}}, \bibinfo {author} {\bibfnamefont {S.}~\bibnamefont {Sk}},\ and\
  \bibinfo {author} {\bibfnamefont {S.}~\bibnamefont {Sarkar}},\ }\bibfield
  {title} {\bibinfo {title} {{Hairy black holes: Nonexistence of short hairs
  and a bound on the light ring size}},\ }\href
  {https://doi.org/10.1103/PhysRevD.108.L041501} {\bibfield  {journal}
  {\bibinfo  {journal} {Phys. Rev. D}\ }\textbf {\bibinfo {volume} {108}},\
  \bibinfo {pages} {L041501} (\bibinfo {year} {2023})},\ \Eprint
  {https://arxiv.org/abs/2306.14193} {arXiv:2306.14193 [gr-qc]} \BibitemShut
  {NoStop}%
\bibitem [{\citenamefont {Ghosh}\ and\ \citenamefont
  {Singha}(2025)}]{Ghosh:2025igz}%
  \BibitemOpen
  \bibfield  {author} {\bibinfo {author} {\bibfnamefont {R.}~\bibnamefont
  {Ghosh}}\ and\ \bibinfo {author} {\bibfnamefont {C.}~\bibnamefont {Singha}},\
  }\bibfield  {title} {\bibinfo {title} {{Can Rotating Black Holes Have Short
  Hairs?}},\ }\href@noop {} {\  (\bibinfo {year} {2025})},\ \Eprint
  {https://arxiv.org/abs/2501.08767} {arXiv:2501.08767 [gr-qc]} \BibitemShut
  {NoStop}%
\bibitem [{\citenamefont {Chrusciel}\ \emph {et~al.}(2012)\citenamefont
  {Chrusciel}, \citenamefont {Lopes~Costa},\ and\ \citenamefont
  {Heusler}}]{Chrusciel:2012jk}%
  \BibitemOpen
  \bibfield  {author} {\bibinfo {author} {\bibfnamefont {P.~T.}\ \bibnamefont
  {Chrusciel}}, \bibinfo {author} {\bibfnamefont {J.}~\bibnamefont
  {Lopes~Costa}},\ and\ \bibinfo {author} {\bibfnamefont {M.}~\bibnamefont
  {Heusler}},\ }\bibfield  {title} {\bibinfo {title} {{Stationary Black Holes:
  Uniqueness and Beyond}},\ }\href {https://doi.org/10.12942/lrr-2012-7}
  {\bibfield  {journal} {\bibinfo  {journal} {Living Rev. Rel.}\ }\textbf
  {\bibinfo {volume} {15}},\ \bibinfo {pages} {7} (\bibinfo {year} {2012})},\
  \Eprint {https://arxiv.org/abs/1205.6112} {arXiv:1205.6112 [gr-qc]}
  \BibitemShut {NoStop}%
\bibitem [{\citenamefont {Herdeiro}\ and\ \citenamefont
  {Radu}(2015)}]{Herdeiro:2015waa}%
  \BibitemOpen
  \bibfield  {author} {\bibinfo {author} {\bibfnamefont {C.~A.~R.}\
  \bibnamefont {Herdeiro}}\ and\ \bibinfo {author} {\bibfnamefont
  {E.}~\bibnamefont {Radu}},\ }\bibfield  {title} {\bibinfo {title}
  {{Asymptotically flat black holes with scalar hair: a review}},\ }\href
  {https://doi.org/10.1142/S0218271815420146} {\bibfield  {journal} {\bibinfo
  {journal} {Int. J. Mod. Phys. D}\ }\textbf {\bibinfo {volume} {24}},\
  \bibinfo {pages} {1542014} (\bibinfo {year} {2015})},\ \Eprint
  {https://arxiv.org/abs/1504.08209} {arXiv:1504.08209 [gr-qc]} \BibitemShut
  {NoStop}%
\bibitem [{\citenamefont {Bizon}(1990)}]{Bizon:1990sr}%
  \BibitemOpen
  \bibfield  {author} {\bibinfo {author} {\bibfnamefont {P.}~\bibnamefont
  {Bizon}},\ }\bibfield  {title} {\bibinfo {title} {{Colored black holes}},\
  }\href {https://doi.org/10.1103/PhysRevLett.64.2844} {\bibfield  {journal}
  {\bibinfo  {journal} {Phys. Rev. Lett.}\ }\textbf {\bibinfo {volume} {64}},\
  \bibinfo {pages} {2844} (\bibinfo {year} {1990})}\BibitemShut {NoStop}%
\bibitem [{\citenamefont {Droz}\ \emph {et~al.}(1991)\citenamefont {Droz},
  \citenamefont {Heusler},\ and\ \citenamefont {Straumann}}]{Droz:1991cx}%
  \BibitemOpen
  \bibfield  {author} {\bibinfo {author} {\bibfnamefont {S.}~\bibnamefont
  {Droz}}, \bibinfo {author} {\bibfnamefont {M.}~\bibnamefont {Heusler}},\ and\
  \bibinfo {author} {\bibfnamefont {N.}~\bibnamefont {Straumann}},\ }\bibfield
  {title} {\bibinfo {title} {{New black hole solutions with hair}},\ }\href
  {https://doi.org/10.1016/0370-2693(91)91592-J} {\bibfield  {journal}
  {\bibinfo  {journal} {Phys. Lett. B}\ }\textbf {\bibinfo {volume} {268}},\
  \bibinfo {pages} {371} (\bibinfo {year} {1991})}\BibitemShut {NoStop}%
\bibitem [{\citenamefont {Breitenlohner}\ \emph {et~al.}(1992)\citenamefont
  {Breitenlohner}, \citenamefont {Forgacs},\ and\ \citenamefont
  {Maison}}]{Breitenlohner:1991aa}%
  \BibitemOpen
  \bibfield  {author} {\bibinfo {author} {\bibfnamefont {P.}~\bibnamefont
  {Breitenlohner}}, \bibinfo {author} {\bibfnamefont {P.}~\bibnamefont
  {Forgacs}},\ and\ \bibinfo {author} {\bibfnamefont {D.}~\bibnamefont
  {Maison}},\ }\bibfield  {title} {\bibinfo {title} {{Gravitating monopole
  solutions}},\ }\href {https://doi.org/10.1016/0550-3213(92)90682-2}
  {\bibfield  {journal} {\bibinfo  {journal} {Nucl. Phys. B}\ }\textbf
  {\bibinfo {volume} {383}},\ \bibinfo {pages} {357} (\bibinfo {year}
  {1992})}\BibitemShut {NoStop}%
\bibitem [{\citenamefont {Greene}\ \emph {et~al.}(1993)\citenamefont {Greene},
  \citenamefont {Mathur},\ and\ \citenamefont {O'Neill}}]{Greene:1992fw}%
  \BibitemOpen
  \bibfield  {author} {\bibinfo {author} {\bibfnamefont {B.~R.}\ \bibnamefont
  {Greene}}, \bibinfo {author} {\bibfnamefont {S.~D.}\ \bibnamefont {Mathur}},\
  and\ \bibinfo {author} {\bibfnamefont {C.~M.}\ \bibnamefont {O'Neill}},\
  }\bibfield  {title} {\bibinfo {title} {{Eluding the no hair conjecture: Black
  holes in spontaneously broken gauge theories}},\ }\href
  {https://doi.org/10.1103/PhysRevD.47.2242} {\bibfield  {journal} {\bibinfo
  {journal} {Phys. Rev. D}\ }\textbf {\bibinfo {volume} {47}},\ \bibinfo
  {pages} {2242} (\bibinfo {year} {1993})},\ \Eprint
  {https://arxiv.org/abs/hep-th/9211007} {arXiv:hep-th/9211007} \BibitemShut
  {NoStop}%
\bibitem [{\citenamefont {Hartmann}\ \emph {et~al.}(2002)\citenamefont
  {Hartmann}, \citenamefont {Kleihaus},\ and\ \citenamefont
  {Kunz}}]{Hartmann:2001ic}%
  \BibitemOpen
  \bibfield  {author} {\bibinfo {author} {\bibfnamefont {B.}~\bibnamefont
  {Hartmann}}, \bibinfo {author} {\bibfnamefont {B.}~\bibnamefont {Kleihaus}},\
  and\ \bibinfo {author} {\bibfnamefont {J.}~\bibnamefont {Kunz}},\ }\bibfield
  {title} {\bibinfo {title} {{Axially symmetric monopoles and black holes in
  Einstein-Yang-Mills-Higgs theory}},\ }\href
  {https://doi.org/10.1103/PhysRevD.65.024027} {\bibfield  {journal} {\bibinfo
  {journal} {Phys. Rev. D}\ }\textbf {\bibinfo {volume} {65}},\ \bibinfo
  {pages} {024027} (\bibinfo {year} {2002})},\ \Eprint
  {https://arxiv.org/abs/hep-th/0108129} {arXiv:hep-th/0108129} \BibitemShut
  {NoStop}%
\bibitem [{\citenamefont {Brito}\ \emph {et~al.}(2013)\citenamefont {Brito},
  \citenamefont {Cardoso},\ and\ \citenamefont {Pani}}]{Brito:2013xaa}%
  \BibitemOpen
  \bibfield  {author} {\bibinfo {author} {\bibfnamefont {R.}~\bibnamefont
  {Brito}}, \bibinfo {author} {\bibfnamefont {V.}~\bibnamefont {Cardoso}},\
  and\ \bibinfo {author} {\bibfnamefont {P.}~\bibnamefont {Pani}},\ }\bibfield
  {title} {\bibinfo {title} {{Black holes with massive graviton hair}},\ }\href
  {https://doi.org/10.1103/PhysRevD.88.064006} {\bibfield  {journal} {\bibinfo
  {journal} {Phys. Rev. D}\ }\textbf {\bibinfo {volume} {88}},\ \bibinfo
  {pages} {064006} (\bibinfo {year} {2013})},\ \Eprint
  {https://arxiv.org/abs/1309.0818} {arXiv:1309.0818 [gr-qc]} \BibitemShut
  {NoStop}%
\bibitem [{\citenamefont {Lavrelashvili}\ and\ \citenamefont
  {Maison}(1993)}]{Lavrelashvili:1992ia}%
  \BibitemOpen
  \bibfield  {author} {\bibinfo {author} {\bibfnamefont {G.~V.}\ \bibnamefont
  {Lavrelashvili}}\ and\ \bibinfo {author} {\bibfnamefont {D.}~\bibnamefont
  {Maison}},\ }\bibfield  {title} {\bibinfo {title} {{Regular and black hole
  solutions of Einstein Yang-Mills Dilaton theory}},\ }\href
  {https://doi.org/10.1016/0550-3213(93)90441-Q} {\bibfield  {journal}
  {\bibinfo  {journal} {Nucl. Phys. B}\ }\textbf {\bibinfo {volume} {410}},\
  \bibinfo {pages} {407} (\bibinfo {year} {1993})}\BibitemShut {NoStop}%
\bibitem [{\citenamefont {Kanti}\ \emph {et~al.}(1996)\citenamefont {Kanti},
  \citenamefont {Mavromatos}, \citenamefont {Rizos}, \citenamefont {Tamvakis},\
  and\ \citenamefont {Winstanley}}]{Kanti:1995vq}%
  \BibitemOpen
  \bibfield  {author} {\bibinfo {author} {\bibfnamefont {P.}~\bibnamefont
  {Kanti}}, \bibinfo {author} {\bibfnamefont {N.~E.}\ \bibnamefont
  {Mavromatos}}, \bibinfo {author} {\bibfnamefont {J.}~\bibnamefont {Rizos}},
  \bibinfo {author} {\bibfnamefont {K.}~\bibnamefont {Tamvakis}},\ and\
  \bibinfo {author} {\bibfnamefont {E.}~\bibnamefont {Winstanley}},\ }\bibfield
   {title} {\bibinfo {title} {{Dilatonic black holes in higher curvature string
  gravity}},\ }\href {https://doi.org/10.1103/PhysRevD.54.5049} {\bibfield
  {journal} {\bibinfo  {journal} {Phys. Rev. D}\ }\textbf {\bibinfo {volume}
  {54}},\ \bibinfo {pages} {5049} (\bibinfo {year} {1996})},\ \Eprint
  {https://arxiv.org/abs/hep-th/9511071} {arXiv:hep-th/9511071} \BibitemShut
  {NoStop}%
\bibitem [{\citenamefont {Lee}\ \emph {et~al.}(2019)\citenamefont {Lee},
  \citenamefont {Lee},\ and\ \citenamefont {Ro}}]{Lee:2018zym}%
  \BibitemOpen
  \bibfield  {author} {\bibinfo {author} {\bibfnamefont {B.-H.}\ \bibnamefont
  {Lee}}, \bibinfo {author} {\bibfnamefont {W.}~\bibnamefont {Lee}},\ and\
  \bibinfo {author} {\bibfnamefont {D.}~\bibnamefont {Ro}},\ }\bibfield
  {title} {\bibinfo {title} {{Expanded evasion of the black hole no-hair
  theorem in dilatonic Einstein-Gauss-Bonnet theory}},\ }\href
  {https://doi.org/10.1103/PhysRevD.99.024002} {\bibfield  {journal} {\bibinfo
  {journal} {Phys. Rev. D}\ }\textbf {\bibinfo {volume} {99}},\ \bibinfo
  {pages} {024002} (\bibinfo {year} {2019})},\ \Eprint
  {https://arxiv.org/abs/1809.05653} {arXiv:1809.05653 [gr-qc]} \BibitemShut
  {NoStop}%
\bibitem [{\citenamefont {Sotiriou}\ and\ \citenamefont
  {Zhou}(2014)}]{Sotiriou:2014pfa}%
  \BibitemOpen
  \bibfield  {author} {\bibinfo {author} {\bibfnamefont {T.~P.}\ \bibnamefont
  {Sotiriou}}\ and\ \bibinfo {author} {\bibfnamefont {S.-Y.}\ \bibnamefont
  {Zhou}},\ }\bibfield  {title} {\bibinfo {title} {{Black hole hair in
  generalized scalar-tensor gravity: An explicit example}},\ }\href
  {https://doi.org/10.1103/PhysRevD.90.124063} {\bibfield  {journal} {\bibinfo
  {journal} {Phys. Rev. D}\ }\textbf {\bibinfo {volume} {90}},\ \bibinfo
  {pages} {124063} (\bibinfo {year} {2014})},\ \Eprint
  {https://arxiv.org/abs/1408.1698} {arXiv:1408.1698 [gr-qc]} \BibitemShut
  {NoStop}%
\bibitem [{\citenamefont {Antoniou}\ \emph {et~al.}(2018)\citenamefont
  {Antoniou}, \citenamefont {Bakopoulos},\ and\ \citenamefont
  {Kanti}}]{Antoniou:2017acq}%
  \BibitemOpen
  \bibfield  {author} {\bibinfo {author} {\bibfnamefont {G.}~\bibnamefont
  {Antoniou}}, \bibinfo {author} {\bibfnamefont {A.}~\bibnamefont
  {Bakopoulos}},\ and\ \bibinfo {author} {\bibfnamefont {P.}~\bibnamefont
  {Kanti}},\ }\bibfield  {title} {\bibinfo {title} {{Evasion of No-Hair
  Theorems and Novel Black-Hole Solutions in Gauss-Bonnet Theories}},\ }\href
  {https://doi.org/10.1103/PhysRevLett.120.131102} {\bibfield  {journal}
  {\bibinfo  {journal} {Phys. Rev. Lett.}\ }\textbf {\bibinfo {volume} {120}},\
  \bibinfo {pages} {131102} (\bibinfo {year} {2018})},\ \Eprint
  {https://arxiv.org/abs/1711.03390} {arXiv:1711.03390 [hep-th]} \BibitemShut
  {NoStop}%
\bibitem [{\citenamefont {Silva}\ \emph {et~al.}(2018)\citenamefont {Silva},
  \citenamefont {Sakstein}, \citenamefont {Gualtieri}, \citenamefont
  {Sotiriou},\ and\ \citenamefont {Berti}}]{Silva:2017uqg}%
  \BibitemOpen
  \bibfield  {author} {\bibinfo {author} {\bibfnamefont {H.~O.}\ \bibnamefont
  {Silva}}, \bibinfo {author} {\bibfnamefont {J.}~\bibnamefont {Sakstein}},
  \bibinfo {author} {\bibfnamefont {L.}~\bibnamefont {Gualtieri}}, \bibinfo
  {author} {\bibfnamefont {T.~P.}\ \bibnamefont {Sotiriou}},\ and\ \bibinfo
  {author} {\bibfnamefont {E.}~\bibnamefont {Berti}},\ }\bibfield  {title}
  {\bibinfo {title} {{Spontaneous scalarization of black holes and compact
  stars from a Gauss-Bonnet coupling}},\ }\href
  {https://doi.org/10.1103/PhysRevLett.120.131104} {\bibfield  {journal}
  {\bibinfo  {journal} {Phys. Rev. Lett.}\ }\textbf {\bibinfo {volume} {120}},\
  \bibinfo {pages} {131104} (\bibinfo {year} {2018})},\ \Eprint
  {https://arxiv.org/abs/1711.02080} {arXiv:1711.02080 [gr-qc]} \BibitemShut
  {NoStop}%
\bibitem [{\citenamefont {Doneva}\ and\ \citenamefont
  {Yazadjiev}(2018)}]{Doneva:2017bvd}%
  \BibitemOpen
  \bibfield  {author} {\bibinfo {author} {\bibfnamefont {D.~D.}\ \bibnamefont
  {Doneva}}\ and\ \bibinfo {author} {\bibfnamefont {S.~S.}\ \bibnamefont
  {Yazadjiev}},\ }\bibfield  {title} {\bibinfo {title} {{New Gauss-Bonnet Black
  Holes with Curvature-Induced Scalarization in Extended Scalar-Tensor
  Theories}},\ }\href {https://doi.org/10.1103/PhysRevLett.120.131103}
  {\bibfield  {journal} {\bibinfo  {journal} {Phys. Rev. Lett.}\ }\textbf
  {\bibinfo {volume} {120}},\ \bibinfo {pages} {131103} (\bibinfo {year}
  {2018})},\ \Eprint {https://arxiv.org/abs/1711.01187} {arXiv:1711.01187
  [gr-qc]} \BibitemShut {NoStop}%
\bibitem [{\citenamefont {Doneva}\ and\ \citenamefont
  {Yazadjiev}(2022)}]{Doneva:2021tvn}%
  \BibitemOpen
  \bibfield  {author} {\bibinfo {author} {\bibfnamefont {D.~D.}\ \bibnamefont
  {Doneva}}\ and\ \bibinfo {author} {\bibfnamefont {S.~S.}\ \bibnamefont
  {Yazadjiev}},\ }\bibfield  {title} {\bibinfo {title} {{Beyond the spontaneous
  scalarization: New fully nonlinear mechanism for the formation of scalarized
  black holes and its dynamical development}},\ }\href
  {https://doi.org/10.1103/PhysRevD.105.L041502} {\bibfield  {journal}
  {\bibinfo  {journal} {Phys. Rev. D}\ }\textbf {\bibinfo {volume} {105}},\
  \bibinfo {pages} {L041502} (\bibinfo {year} {2022})},\ \Eprint
  {https://arxiv.org/abs/2107.01738} {arXiv:2107.01738 [gr-qc]} \BibitemShut
  {NoStop}%
\bibitem [{\citenamefont {Macedo}\ \emph {et~al.}(2019)\citenamefont {Macedo},
  \citenamefont {Sakstein}, \citenamefont {Berti}, \citenamefont {Gualtieri},
  \citenamefont {Silva},\ and\ \citenamefont {Sotiriou}}]{Macedo:2019sem}%
  \BibitemOpen
  \bibfield  {author} {\bibinfo {author} {\bibfnamefont {C.~F.~B.}\
  \bibnamefont {Macedo}}, \bibinfo {author} {\bibfnamefont {J.}~\bibnamefont
  {Sakstein}}, \bibinfo {author} {\bibfnamefont {E.}~\bibnamefont {Berti}},
  \bibinfo {author} {\bibfnamefont {L.}~\bibnamefont {Gualtieri}}, \bibinfo
  {author} {\bibfnamefont {H.~O.}\ \bibnamefont {Silva}},\ and\ \bibinfo
  {author} {\bibfnamefont {T.~P.}\ \bibnamefont {Sotiriou}},\ }\bibfield
  {title} {\bibinfo {title} {{Self-interactions and Spontaneous Black Hole
  Scalarization}},\ }\href {https://doi.org/10.1103/PhysRevD.99.104041}
  {\bibfield  {journal} {\bibinfo  {journal} {Phys. Rev. D}\ }\textbf {\bibinfo
  {volume} {99}},\ \bibinfo {pages} {104041} (\bibinfo {year} {2019})},\
  \Eprint {https://arxiv.org/abs/1903.06784} {arXiv:1903.06784 [gr-qc]}
  \BibitemShut {NoStop}%
\bibitem [{\citenamefont {Kuan}\ \emph {et~al.}(2021)\citenamefont {Kuan},
  \citenamefont {Doneva},\ and\ \citenamefont {Yazadjiev}}]{Kuan:2021lol}%
  \BibitemOpen
  \bibfield  {author} {\bibinfo {author} {\bibfnamefont {H.-J.}\ \bibnamefont
  {Kuan}}, \bibinfo {author} {\bibfnamefont {D.~D.}\ \bibnamefont {Doneva}},\
  and\ \bibinfo {author} {\bibfnamefont {S.~S.}\ \bibnamefont {Yazadjiev}},\
  }\bibfield  {title} {\bibinfo {title} {{Dynamical Formation of Scalarized
  Black Holes and Neutron Stars through Stellar Core Collapse}},\ }\href
  {https://doi.org/10.1103/PhysRevLett.127.161103} {\bibfield  {journal}
  {\bibinfo  {journal} {Phys. Rev. Lett.}\ }\textbf {\bibinfo {volume} {127}},\
  \bibinfo {pages} {161103} (\bibinfo {year} {2021})},\ \Eprint
  {https://arxiv.org/abs/2103.11999} {arXiv:2103.11999 [gr-qc]} \BibitemShut
  {NoStop}%
\bibitem [{\citenamefont {Dima}\ \emph {et~al.}(2020)\citenamefont {Dima},
  \citenamefont {Barausse}, \citenamefont {Franchini},\ and\ \citenamefont
  {Sotiriou}}]{Dima:2020yac}%
  \BibitemOpen
  \bibfield  {author} {\bibinfo {author} {\bibfnamefont {A.}~\bibnamefont
  {Dima}}, \bibinfo {author} {\bibfnamefont {E.}~\bibnamefont {Barausse}},
  \bibinfo {author} {\bibfnamefont {N.}~\bibnamefont {Franchini}},\ and\
  \bibinfo {author} {\bibfnamefont {T.~P.}\ \bibnamefont {Sotiriou}},\
  }\bibfield  {title} {\bibinfo {title} {{Spin-induced black hole spontaneous
  scalarization}},\ }\href {https://doi.org/10.1103/PhysRevLett.125.231101}
  {\bibfield  {journal} {\bibinfo  {journal} {Phys. Rev. Lett.}\ }\textbf
  {\bibinfo {volume} {125}},\ \bibinfo {pages} {231101} (\bibinfo {year}
  {2020})},\ \Eprint {https://arxiv.org/abs/2006.03095} {arXiv:2006.03095
  [gr-qc]} \BibitemShut {NoStop}%
\bibitem [{\citenamefont {Berti}\ \emph {et~al.}(2021)\citenamefont {Berti},
  \citenamefont {Collodel}, \citenamefont {Kleihaus},\ and\ \citenamefont
  {Kunz}}]{Berti:2020kgk}%
  \BibitemOpen
  \bibfield  {author} {\bibinfo {author} {\bibfnamefont {E.}~\bibnamefont
  {Berti}}, \bibinfo {author} {\bibfnamefont {L.~G.}\ \bibnamefont {Collodel}},
  \bibinfo {author} {\bibfnamefont {B.}~\bibnamefont {Kleihaus}},\ and\
  \bibinfo {author} {\bibfnamefont {J.}~\bibnamefont {Kunz}},\ }\bibfield
  {title} {\bibinfo {title} {{Spin-induced black-hole scalarization in
  Einstein-scalar-Gauss-Bonnet theory}},\ }\href
  {https://doi.org/10.1103/PhysRevLett.126.011104} {\bibfield  {journal}
  {\bibinfo  {journal} {Phys. Rev. Lett.}\ }\textbf {\bibinfo {volume} {126}},\
  \bibinfo {pages} {011104} (\bibinfo {year} {2021})},\ \Eprint
  {https://arxiv.org/abs/2009.03905} {arXiv:2009.03905 [gr-qc]} \BibitemShut
  {NoStop}%
\bibitem [{\citenamefont {Herdeiro}\ \emph {et~al.}(2021)\citenamefont
  {Herdeiro}, \citenamefont {Radu}, \citenamefont {Silva}, \citenamefont
  {Sotiriou},\ and\ \citenamefont {Yunes}}]{Herdeiro:2020wei}%
  \BibitemOpen
  \bibfield  {author} {\bibinfo {author} {\bibfnamefont {C.~A.~R.}\
  \bibnamefont {Herdeiro}}, \bibinfo {author} {\bibfnamefont {E.}~\bibnamefont
  {Radu}}, \bibinfo {author} {\bibfnamefont {H.~O.}\ \bibnamefont {Silva}},
  \bibinfo {author} {\bibfnamefont {T.~P.}\ \bibnamefont {Sotiriou}},\ and\
  \bibinfo {author} {\bibfnamefont {N.}~\bibnamefont {Yunes}},\ }\bibfield
  {title} {\bibinfo {title} {{Spin-induced scalarized black holes}},\ }\href
  {https://doi.org/10.1103/PhysRevLett.126.011103} {\bibfield  {journal}
  {\bibinfo  {journal} {Phys. Rev. Lett.}\ }\textbf {\bibinfo {volume} {126}},\
  \bibinfo {pages} {011103} (\bibinfo {year} {2021})},\ \Eprint
  {https://arxiv.org/abs/2009.03904} {arXiv:2009.03904 [gr-qc]} \BibitemShut
  {NoStop}%
\bibitem [{\citenamefont {Cunha}\ \emph {et~al.}(2019)\citenamefont {Cunha},
  \citenamefont {Herdeiro},\ and\ \citenamefont {Radu}}]{Cunha:2019dwb}%
  \BibitemOpen
  \bibfield  {author} {\bibinfo {author} {\bibfnamefont {P.~V.~P.}\
  \bibnamefont {Cunha}}, \bibinfo {author} {\bibfnamefont {C.~A.~R.}\
  \bibnamefont {Herdeiro}},\ and\ \bibinfo {author} {\bibfnamefont
  {E.}~\bibnamefont {Radu}},\ }\bibfield  {title} {\bibinfo {title}
  {{Spontaneously Scalarized Kerr Black Holes in Extended
  Scalar-Tensor\textendash{}Gauss-Bonnet Gravity}},\ }\href
  {https://doi.org/10.1103/PhysRevLett.123.011101} {\bibfield  {journal}
  {\bibinfo  {journal} {Phys. Rev. Lett.}\ }\textbf {\bibinfo {volume} {123}},\
  \bibinfo {pages} {011101} (\bibinfo {year} {2019})},\ \Eprint
  {https://arxiv.org/abs/1904.09997} {arXiv:1904.09997 [gr-qc]} \BibitemShut
  {NoStop}%
\bibitem [{\citenamefont {Hod}(2020)}]{Hod:2020jjy}%
  \BibitemOpen
  \bibfield  {author} {\bibinfo {author} {\bibfnamefont {S.}~\bibnamefont
  {Hod}},\ }\bibfield  {title} {\bibinfo {title} {{Onset of spontaneous
  scalarization in spinning Gauss-Bonnet black holes}},\ }\href
  {https://doi.org/10.1103/PhysRevD.102.084060} {\bibfield  {journal} {\bibinfo
   {journal} {Phys. Rev. D}\ }\textbf {\bibinfo {volume} {102}},\ \bibinfo
  {pages} {084060} (\bibinfo {year} {2020})},\ \Eprint
  {https://arxiv.org/abs/2006.09399} {arXiv:2006.09399 [gr-qc]} \BibitemShut
  {NoStop}%
\bibitem [{\citenamefont {Herdeiro}\ and\ \citenamefont
  {Radu}(2014)}]{Herdeiro:2014goa}%
  \BibitemOpen
  \bibfield  {author} {\bibinfo {author} {\bibfnamefont {C.~A.~R.}\
  \bibnamefont {Herdeiro}}\ and\ \bibinfo {author} {\bibfnamefont
  {E.}~\bibnamefont {Radu}},\ }\bibfield  {title} {\bibinfo {title} {{Kerr
  black holes with scalar hair}},\ }\href
  {https://doi.org/10.1103/PhysRevLett.112.221101} {\bibfield  {journal}
  {\bibinfo  {journal} {Phys. Rev. Lett.}\ }\textbf {\bibinfo {volume} {112}},\
  \bibinfo {pages} {221101} (\bibinfo {year} {2014})},\ \Eprint
  {https://arxiv.org/abs/1403.2757} {arXiv:1403.2757 [gr-qc]} \BibitemShut
  {NoStop}%
\bibitem [{\citenamefont {Herdeiro}\ \emph {et~al.}(2016)\citenamefont
  {Herdeiro}, \citenamefont {Radu},\ and\ \citenamefont
  {R\'unarsson}}]{Herdeiro:2016tmi}%
  \BibitemOpen
  \bibfield  {author} {\bibinfo {author} {\bibfnamefont {C.}~\bibnamefont
  {Herdeiro}}, \bibinfo {author} {\bibfnamefont {E.}~\bibnamefont {Radu}},\
  and\ \bibinfo {author} {\bibfnamefont {H.}~\bibnamefont {R\'unarsson}},\
  }\bibfield  {title} {\bibinfo {title} {{Kerr black holes with Proca hair}},\
  }\href {https://doi.org/10.1088/0264-9381/33/15/154001} {\bibfield  {journal}
  {\bibinfo  {journal} {Class. Quant. Grav.}\ }\textbf {\bibinfo {volume}
  {33}},\ \bibinfo {pages} {154001} (\bibinfo {year} {2016})},\ \Eprint
  {https://arxiv.org/abs/1603.02687} {arXiv:1603.02687 [gr-qc]} \BibitemShut
  {NoStop}%
\bibitem [{\citenamefont {Santos}\ \emph {et~al.}(2020)\citenamefont {Santos},
  \citenamefont {Benone}, \citenamefont {Crispino}, \citenamefont {Herdeiro},\
  and\ \citenamefont {Radu}}]{Santos:2020pmh}%
  \BibitemOpen
  \bibfield  {author} {\bibinfo {author} {\bibfnamefont {N.~M.}\ \bibnamefont
  {Santos}}, \bibinfo {author} {\bibfnamefont {C.~L.}\ \bibnamefont {Benone}},
  \bibinfo {author} {\bibfnamefont {L.~C.~B.}\ \bibnamefont {Crispino}},
  \bibinfo {author} {\bibfnamefont {C.~A.~R.}\ \bibnamefont {Herdeiro}},\ and\
  \bibinfo {author} {\bibfnamefont {E.}~\bibnamefont {Radu}},\ }\bibfield
  {title} {\bibinfo {title} {{Black holes with synchronised Proca hair: linear
  clouds and fundamental non-linear solutions}},\ }\href
  {https://doi.org/10.1007/JHEP07(2020)010} {\bibfield  {journal} {\bibinfo
  {journal} {JHEP}\ }\textbf {\bibinfo {volume} {07}},\ \bibinfo {pages}
  {010}},\ \Eprint {https://arxiv.org/abs/2004.09536} {arXiv:2004.09536
  [gr-qc]} \BibitemShut {NoStop}%
\bibitem [{\citenamefont {Corichi}\ \emph {et~al.}(2006)\citenamefont
  {Corichi}, \citenamefont {Nucamendi},\ and\ \citenamefont
  {Salgado}}]{Corichi:2005pa}%
  \BibitemOpen
  \bibfield  {author} {\bibinfo {author} {\bibfnamefont {A.}~\bibnamefont
  {Corichi}}, \bibinfo {author} {\bibfnamefont {U.}~\bibnamefont {Nucamendi}},\
  and\ \bibinfo {author} {\bibfnamefont {M.}~\bibnamefont {Salgado}},\
  }\bibfield  {title} {\bibinfo {title} {{Scalar hairy black holes and
  scalarons in the isolated horizons formalism}},\ }\href
  {https://doi.org/10.1103/PhysRevD.73.084002} {\bibfield  {journal} {\bibinfo
  {journal} {Phys. Rev. D}\ }\textbf {\bibinfo {volume} {73}},\ \bibinfo
  {pages} {084002} (\bibinfo {year} {2006})},\ \Eprint
  {https://arxiv.org/abs/gr-qc/0504126} {arXiv:gr-qc/0504126} \BibitemShut
  {NoStop}%
\bibitem [{\citenamefont {Feng}\ \emph {et~al.}(2014)\citenamefont {Feng},
  \citenamefont {Lu},\ and\ \citenamefont {Wen}}]{Feng:2013tza}%
  \BibitemOpen
  \bibfield  {author} {\bibinfo {author} {\bibfnamefont {X.-H.}\ \bibnamefont
  {Feng}}, \bibinfo {author} {\bibfnamefont {H.}~\bibnamefont {Lu}},\ and\
  \bibinfo {author} {\bibfnamefont {Q.}~\bibnamefont {Wen}},\ }\bibfield
  {title} {\bibinfo {title} {{Scalar Hairy Black Holes in General
  Dimensions}},\ }\href {https://doi.org/10.1103/PhysRevD.89.044014} {\bibfield
   {journal} {\bibinfo  {journal} {Phys. Rev. D}\ }\textbf {\bibinfo {volume}
  {89}},\ \bibinfo {pages} {044014} (\bibinfo {year} {2014})},\ \Eprint
  {https://arxiv.org/abs/1312.5374} {arXiv:1312.5374 [hep-th]} \BibitemShut
  {NoStop}%
\bibitem [{\citenamefont {Chew}\ \emph {et~al.}(2023)\citenamefont {Chew},
  \citenamefont {Yeom},\ and\ \citenamefont
  {Bl\'azquez-Salcedo}}]{Chew:2022enh}%
  \BibitemOpen
  \bibfield  {author} {\bibinfo {author} {\bibfnamefont {X.~Y.}\ \bibnamefont
  {Chew}}, \bibinfo {author} {\bibfnamefont {D.-h.}\ \bibnamefont {Yeom}},\
  and\ \bibinfo {author} {\bibfnamefont {J.~L.}\ \bibnamefont
  {Bl\'azquez-Salcedo}},\ }\bibfield  {title} {\bibinfo {title} {{Properties of
  scalar hairy black holes and scalarons with asymmetric potential}},\ }\href
  {https://doi.org/10.1103/PhysRevD.108.044020} {\bibfield  {journal} {\bibinfo
   {journal} {Phys. Rev. D}\ }\textbf {\bibinfo {volume} {108}},\ \bibinfo
  {pages} {044020} (\bibinfo {year} {2023})},\ \Eprint
  {https://arxiv.org/abs/2210.01313} {arXiv:2210.01313 [gr-qc]} \BibitemShut
  {NoStop}%
\bibitem [{\citenamefont {Chew}\ and\ \citenamefont
  {Yeom}(2024)}]{Chew:2024rin}%
  \BibitemOpen
  \bibfield  {author} {\bibinfo {author} {\bibfnamefont {X.~Y.}\ \bibnamefont
  {Chew}}\ and\ \bibinfo {author} {\bibfnamefont {D.-h.}\ \bibnamefont
  {Yeom}},\ }\bibfield  {title} {\bibinfo {title} {{Hairy Reissner-Nordstr\"om
  black holes with asymmetric vacua}},\ }\href
  {https://doi.org/10.1103/PhysRevD.110.044036} {\bibfield  {journal} {\bibinfo
   {journal} {Phys. Rev. D}\ }\textbf {\bibinfo {volume} {110}},\ \bibinfo
  {pages} {044036} (\bibinfo {year} {2024})},\ \Eprint
  {https://arxiv.org/abs/2401.09039} {arXiv:2401.09039 [gr-qc]} \BibitemShut
  {NoStop}%
\bibitem [{\citenamefont {Chew}\ and\ \citenamefont
  {Lim}(2024)}]{Chew:2023olq}%
  \BibitemOpen
  \bibfield  {author} {\bibinfo {author} {\bibfnamefont {X.~Y.}\ \bibnamefont
  {Chew}}\ and\ \bibinfo {author} {\bibfnamefont {K.-G.}\ \bibnamefont {Lim}},\
  }\bibfield  {title} {\bibinfo {title} {{Scalar hairy black holes with an
  inverted Mexican-hat potential}},\ }\href
  {https://doi.org/10.1103/PhysRevD.109.064039} {\bibfield  {journal} {\bibinfo
   {journal} {Phys. Rev. D}\ }\textbf {\bibinfo {volume} {109}},\ \bibinfo
  {pages} {064039} (\bibinfo {year} {2024})},\ \Eprint
  {https://arxiv.org/abs/2307.13972} {arXiv:2307.13972 [gr-qc]} \BibitemShut
  {NoStop}%
\bibitem [{\citenamefont {Chew}\ and\ \citenamefont
  {Myung}(2024)}]{Chew:2024evh}%
  \BibitemOpen
  \bibfield  {author} {\bibinfo {author} {\bibfnamefont {X.~Y.}\ \bibnamefont
  {Chew}}\ and\ \bibinfo {author} {\bibfnamefont {Y.~S.}\ \bibnamefont
  {Myung}},\ }\bibfield  {title} {\bibinfo {title} {{Simplest model of a
  scalarized black hole in the Einstein-Klein-Gordon theory}},\ }\href
  {https://doi.org/10.1103/PhysRevD.110.044011} {\bibfield  {journal} {\bibinfo
   {journal} {Phys. Rev. D}\ }\textbf {\bibinfo {volume} {110}},\ \bibinfo
  {pages} {044011} (\bibinfo {year} {2024})},\ \Eprint
  {https://arxiv.org/abs/2405.04921} {arXiv:2405.04921 [gr-qc]} \BibitemShut
  {NoStop}%
\bibitem [{\citenamefont {Herdeiro}\ \emph {et~al.}(2018)\citenamefont
  {Herdeiro}, \citenamefont {Radu}, \citenamefont {Sanchis-Gual},\ and\
  \citenamefont {Font}}]{Herdeiro:2018wub}%
  \BibitemOpen
  \bibfield  {author} {\bibinfo {author} {\bibfnamefont {C.~A.~R.}\
  \bibnamefont {Herdeiro}}, \bibinfo {author} {\bibfnamefont {E.}~\bibnamefont
  {Radu}}, \bibinfo {author} {\bibfnamefont {N.}~\bibnamefont {Sanchis-Gual}},\
  and\ \bibinfo {author} {\bibfnamefont {J.~A.}\ \bibnamefont {Font}},\
  }\bibfield  {title} {\bibinfo {title} {{Spontaneous Scalarization of Charged
  Black Holes}},\ }\href {https://doi.org/10.1103/PhysRevLett.121.101102}
  {\bibfield  {journal} {\bibinfo  {journal} {Phys. Rev. Lett.}\ }\textbf
  {\bibinfo {volume} {121}},\ \bibinfo {pages} {101102} (\bibinfo {year}
  {2018})},\ \Eprint {https://arxiv.org/abs/1806.05190} {arXiv:1806.05190
  [gr-qc]} \BibitemShut {NoStop}%
\bibitem [{\citenamefont {Fernandes}(2020)}]{Fernandes:2020gay}%
  \BibitemOpen
  \bibfield  {author} {\bibinfo {author} {\bibfnamefont {P.~G.~S.}\
  \bibnamefont {Fernandes}},\ }\bibfield  {title} {\bibinfo {title}
  {{Einstein\textendash{}Maxwell-scalar black holes with massive and
  self-interacting scalar hair}},\ }\href
  {https://doi.org/10.1016/j.dark.2020.100716} {\bibfield  {journal} {\bibinfo
  {journal} {Phys. Dark Univ.}\ }\textbf {\bibinfo {volume} {30}},\ \bibinfo
  {pages} {100716} (\bibinfo {year} {2020})},\ \Eprint
  {https://arxiv.org/abs/2003.01045} {arXiv:2003.01045 [gr-qc]} \BibitemShut
  {NoStop}%
\bibitem [{\citenamefont {Brihaye}\ and\ \citenamefont
  {Hartmann}(2022)}]{Brihaye:2021mqk}%
  \BibitemOpen
  \bibfield  {author} {\bibinfo {author} {\bibfnamefont {Y.}~\bibnamefont
  {Brihaye}}\ and\ \bibinfo {author} {\bibfnamefont {B.}~\bibnamefont
  {Hartmann}},\ }\bibfield  {title} {\bibinfo {title} {{Boson stars and black
  holes with wavy scalar hair}},\ }\href
  {https://doi.org/10.1103/PhysRevD.105.104063} {\bibfield  {journal} {\bibinfo
   {journal} {Phys. Rev. D}\ }\textbf {\bibinfo {volume} {105}},\ \bibinfo
  {pages} {104063} (\bibinfo {year} {2022})},\ \Eprint
  {https://arxiv.org/abs/2112.12830} {arXiv:2112.12830 [gr-qc]} \BibitemShut
  {NoStop}%
\bibitem [{\citenamefont {Hong}\ \emph
  {et~al.}(2020{\natexlab{a}})\citenamefont {Hong}, \citenamefont {Suzuki},\
  and\ \citenamefont {Yamada}}]{Hong:2020miv}%
  \BibitemOpen
  \bibfield  {author} {\bibinfo {author} {\bibfnamefont {J.-P.}\ \bibnamefont
  {Hong}}, \bibinfo {author} {\bibfnamefont {M.}~\bibnamefont {Suzuki}},\ and\
  \bibinfo {author} {\bibfnamefont {M.}~\bibnamefont {Yamada}},\ }\bibfield
  {title} {\bibinfo {title} {{Spherically Symmetric Scalar Hair for Charged
  Black Holes}},\ }\href {https://doi.org/10.1103/PhysRevLett.125.111104}
  {\bibfield  {journal} {\bibinfo  {journal} {Phys. Rev. Lett.}\ }\textbf
  {\bibinfo {volume} {125}},\ \bibinfo {pages} {111104} (\bibinfo {year}
  {2020}{\natexlab{a}})},\ \Eprint {https://arxiv.org/abs/2004.03148}
  {arXiv:2004.03148 [gr-qc]} \BibitemShut {NoStop}%
\bibitem [{\citenamefont {Hong}\ \emph
  {et~al.}(2020{\natexlab{b}})\citenamefont {Hong}, \citenamefont {Suzuki},\
  and\ \citenamefont {Yamada}}]{Hong:2019mcj}%
  \BibitemOpen
  \bibfield  {author} {\bibinfo {author} {\bibfnamefont {J.-P.}\ \bibnamefont
  {Hong}}, \bibinfo {author} {\bibfnamefont {M.}~\bibnamefont {Suzuki}},\ and\
  \bibinfo {author} {\bibfnamefont {M.}~\bibnamefont {Yamada}},\ }\bibfield
  {title} {\bibinfo {title} {{Charged black holes in non-linear Q-clouds with
  O(3) symmetry}},\ }\href {https://doi.org/10.1016/j.physletb.2020.135324}
  {\bibfield  {journal} {\bibinfo  {journal} {Phys. Lett. B}\ }\textbf
  {\bibinfo {volume} {803}},\ \bibinfo {pages} {135324} (\bibinfo {year}
  {2020}{\natexlab{b}})},\ \Eprint {https://arxiv.org/abs/1907.04982}
  {arXiv:1907.04982 [gr-qc]} \BibitemShut {NoStop}%
\bibitem [{\citenamefont {Herdeiro}\ and\ \citenamefont
  {Radu}(2020)}]{Herdeiro:2020xmb}%
  \BibitemOpen
  \bibfield  {author} {\bibinfo {author} {\bibfnamefont {C.~A.~R.}\
  \bibnamefont {Herdeiro}}\ and\ \bibinfo {author} {\bibfnamefont
  {E.}~\bibnamefont {Radu}},\ }\bibfield  {title} {\bibinfo {title} {{Spherical
  electro-vacuum black holes with resonant, scalar $Q$-hair}},\ }\href
  {https://doi.org/10.1140/epjc/s10052-020-7976-9} {\bibfield  {journal}
  {\bibinfo  {journal} {Eur. Phys. J. C}\ }\textbf {\bibinfo {volume} {80}},\
  \bibinfo {pages} {390} (\bibinfo {year} {2020})},\ \Eprint
  {https://arxiv.org/abs/2004.00336} {arXiv:2004.00336 [gr-qc]} \BibitemShut
  {NoStop}%
\bibitem [{\citenamefont {Zhang}\ \emph {et~al.}(2022)\citenamefont {Zhang},
  \citenamefont {Chen}, \citenamefont {Liu}, \citenamefont {Luo}, \citenamefont
  {Tian},\ and\ \citenamefont {Wang}}]{Zhang:2021nnn}%
  \BibitemOpen
  \bibfield  {author} {\bibinfo {author} {\bibfnamefont {C.-Y.}\ \bibnamefont
  {Zhang}}, \bibinfo {author} {\bibfnamefont {Q.}~\bibnamefont {Chen}},
  \bibinfo {author} {\bibfnamefont {Y.}~\bibnamefont {Liu}}, \bibinfo {author}
  {\bibfnamefont {W.-K.}\ \bibnamefont {Luo}}, \bibinfo {author} {\bibfnamefont
  {Y.}~\bibnamefont {Tian}},\ and\ \bibinfo {author} {\bibfnamefont
  {B.}~\bibnamefont {Wang}},\ }\bibfield  {title} {\bibinfo {title} {{Critical
  Phenomena in Dynamical Scalarization of Charged Black Holes}},\ }\href
  {https://doi.org/10.1103/PhysRevLett.128.161105} {\bibfield  {journal}
  {\bibinfo  {journal} {Phys. Rev. Lett.}\ }\textbf {\bibinfo {volume} {128}},\
  \bibinfo {pages} {161105} (\bibinfo {year} {2022})},\ \Eprint
  {https://arxiv.org/abs/2112.07455} {arXiv:2112.07455 [gr-qc]} \BibitemShut
  {NoStop}%
\bibitem [{\citenamefont {Zou}\ and\ \citenamefont
  {Myung}(2019)}]{Zou:2019bpt}%
  \BibitemOpen
  \bibfield  {author} {\bibinfo {author} {\bibfnamefont {D.-C.}\ \bibnamefont
  {Zou}}\ and\ \bibinfo {author} {\bibfnamefont {Y.~S.}\ \bibnamefont
  {Myung}},\ }\bibfield  {title} {\bibinfo {title} {{Scalarized charged black
  holes with scalar mass term}},\ }\href
  {https://doi.org/10.1103/PhysRevD.100.124055} {\bibfield  {journal} {\bibinfo
   {journal} {Phys. Rev. D}\ }\textbf {\bibinfo {volume} {100}},\ \bibinfo
  {pages} {124055} (\bibinfo {year} {2019})},\ \Eprint
  {https://arxiv.org/abs/1909.11859} {arXiv:1909.11859 [gr-qc]} \BibitemShut
  {NoStop}%
\bibitem [{\citenamefont {Barton}\ \emph {et~al.}(2021)\citenamefont {Barton},
  \citenamefont {Hartmann}, \citenamefont {Kleihaus},\ and\ \citenamefont
  {Kunz}}]{Barton:2021wfj}%
  \BibitemOpen
  \bibfield  {author} {\bibinfo {author} {\bibfnamefont {S.}~\bibnamefont
  {Barton}}, \bibinfo {author} {\bibfnamefont {B.}~\bibnamefont {Hartmann}},
  \bibinfo {author} {\bibfnamefont {B.}~\bibnamefont {Kleihaus}},\ and\
  \bibinfo {author} {\bibfnamefont {J.}~\bibnamefont {Kunz}},\ }\bibfield
  {title} {\bibinfo {title} {{Spontaneously vectorized Einstein-Gauss-Bonnet
  black holes}},\ }\href {https://doi.org/10.1016/j.physletb.2021.136336}
  {\bibfield  {journal} {\bibinfo  {journal} {Phys. Lett. B}\ }\textbf
  {\bibinfo {volume} {817}},\ \bibinfo {pages} {136336} (\bibinfo {year}
  {2021})},\ \Eprint {https://arxiv.org/abs/2103.01651} {arXiv:2103.01651
  [gr-qc]} \BibitemShut {NoStop}%
\bibitem [{\citenamefont {Gervalle}\ and\ \citenamefont
  {Volkov}(2024)}]{Gervalle:2024yxj}%
  \BibitemOpen
  \bibfield  {author} {\bibinfo {author} {\bibfnamefont {R.}~\bibnamefont
  {Gervalle}}\ and\ \bibinfo {author} {\bibfnamefont {M.~S.}\ \bibnamefont
  {Volkov}},\ }\bibfield  {title} {\bibinfo {title} {{Black Holes with
  Electroweak Hair}},\ }\href {https://doi.org/10.1103/PhysRevLett.133.171402}
  {\bibfield  {journal} {\bibinfo  {journal} {Phys. Rev. Lett.}\ }\textbf
  {\bibinfo {volume} {133}},\ \bibinfo {pages} {171402} (\bibinfo {year}
  {2024})},\ \Eprint {https://arxiv.org/abs/2406.14357} {arXiv:2406.14357
  [hep-th]} \BibitemShut {NoStop}%
\bibitem [{\citenamefont {Doneva}\ \emph {et~al.}(2024)\citenamefont {Doneva},
  \citenamefont {Ramazano\u{g}lu}, \citenamefont {Silva}, \citenamefont
  {Sotiriou},\ and\ \citenamefont {Yazadjiev}}]{Doneva:2022ewd}%
  \BibitemOpen
  \bibfield  {author} {\bibinfo {author} {\bibfnamefont {D.~D.}\ \bibnamefont
  {Doneva}}, \bibinfo {author} {\bibfnamefont {F.~M.}\ \bibnamefont
  {Ramazano\u{g}lu}}, \bibinfo {author} {\bibfnamefont {H.~O.}\ \bibnamefont
  {Silva}}, \bibinfo {author} {\bibfnamefont {T.~P.}\ \bibnamefont
  {Sotiriou}},\ and\ \bibinfo {author} {\bibfnamefont {S.~S.}\ \bibnamefont
  {Yazadjiev}},\ }\bibfield  {title} {\bibinfo {title} {{Spontaneous
  scalarization}},\ }\href {https://doi.org/10.1103/RevModPhys.96.015004}
  {\bibfield  {journal} {\bibinfo  {journal} {Rev. Mod. Phys.}\ }\textbf
  {\bibinfo {volume} {96}},\ \bibinfo {pages} {015004} (\bibinfo {year}
  {2024})},\ \Eprint {https://arxiv.org/abs/2211.01766} {arXiv:2211.01766
  [gr-qc]} \BibitemShut {NoStop}%
\bibitem [{\citenamefont {Brown}\ and\ \citenamefont
  {Husain}(1997)}]{Brown:1997jv}%
  \BibitemOpen
  \bibfield  {author} {\bibinfo {author} {\bibfnamefont {J.~D.}\ \bibnamefont
  {Brown}}\ and\ \bibinfo {author} {\bibfnamefont {V.}~\bibnamefont {Husain}},\
  }\bibfield  {title} {\bibinfo {title} {{Black holes with short hair}},\
  }\href {https://doi.org/10.1142/S0218271897000340} {\bibfield  {journal}
  {\bibinfo  {journal} {Int. J. Mod. Phys. D}\ }\textbf {\bibinfo {volume}
  {6}},\ \bibinfo {pages} {563} (\bibinfo {year} {1997})},\ \Eprint
  {https://arxiv.org/abs/gr-qc/9707027} {arXiv:gr-qc/9707027} \BibitemShut
  {NoStop}%
\bibitem [{\citenamefont {Nunez}\ \emph {et~al.}(1996)\citenamefont {Nunez},
  \citenamefont {Quevedo},\ and\ \citenamefont {Sudarsky}}]{Nunez:1996xv}%
  \BibitemOpen
  \bibfield  {author} {\bibinfo {author} {\bibfnamefont {D.}~\bibnamefont
  {Nunez}}, \bibinfo {author} {\bibfnamefont {H.}~\bibnamefont {Quevedo}},\
  and\ \bibinfo {author} {\bibfnamefont {D.}~\bibnamefont {Sudarsky}},\
  }\bibfield  {title} {\bibinfo {title} {{Black holes have no short hair}},\
  }\href {https://doi.org/10.1103/PhysRevLett.76.571} {\bibfield  {journal}
  {\bibinfo  {journal} {Phys. Rev. Lett.}\ }\textbf {\bibinfo {volume} {76}},\
  \bibinfo {pages} {571} (\bibinfo {year} {1996})},\ \Eprint
  {https://arxiv.org/abs/gr-qc/9601020} {arXiv:gr-qc/9601020} \BibitemShut
  {NoStop}%
\bibitem [{\citenamefont {Tang}\ and\ \citenamefont {Xu}(2022)}]{Tang:2022uwi}%
  \BibitemOpen
  \bibfield  {author} {\bibinfo {author} {\bibfnamefont {M.}~\bibnamefont
  {Tang}}\ and\ \bibinfo {author} {\bibfnamefont {Z.}~\bibnamefont {Xu}},\
  }\bibfield  {title} {\bibinfo {title} {{The no-hair theorem and black hole
  shadows}},\ }\href {https://doi.org/10.1007/JHEP12(2022)125} {\bibfield
  {journal} {\bibinfo  {journal} {JHEP}\ }\textbf {\bibinfo {volume} {12}},\
  \bibinfo {pages} {125}},\ \Eprint {https://arxiv.org/abs/2209.08202}
  {arXiv:2209.08202 [gr-qc]} \BibitemShut {NoStop}%
\bibitem [{\citenamefont {Zhao}\ \emph
  {et~al.}(2024{\natexlab{a}})\citenamefont {Zhao}, \citenamefont {Tang},\ and\
  \citenamefont {Xu}}]{Zhao:2024qzg}%
  \BibitemOpen
  \bibfield  {author} {\bibinfo {author} {\bibfnamefont {M.}~\bibnamefont
  {Zhao}}, \bibinfo {author} {\bibfnamefont {M.}~\bibnamefont {Tang}},\ and\
  \bibinfo {author} {\bibfnamefont {Z.}~\bibnamefont {Xu}},\ }\bibfield
  {title} {\bibinfo {title} {{Testing the weak cosmic censorship conjecture in
  short haired black holes}},\ }\href
  {https://doi.org/10.1140/epjc/s10052-024-12837-z} {\bibfield  {journal}
  {\bibinfo  {journal} {Eur. Phys. J. C}\ }\textbf {\bibinfo {volume} {84}},\
  \bibinfo {pages} {497} (\bibinfo {year} {2024}{\natexlab{a}})},\ \Eprint
  {https://arxiv.org/abs/2402.16373} {arXiv:2402.16373 [gr-qc]} \BibitemShut
  {NoStop}%
\bibitem [{\citenamefont {Zhao}\ \emph
  {et~al.}(2024{\natexlab{b}})\citenamefont {Zhao}, \citenamefont {Tang},\ and\
  \citenamefont {Xu}}]{Zhao:2024hep}%
  \BibitemOpen
  \bibfield  {author} {\bibinfo {author} {\bibfnamefont {L.}~\bibnamefont
  {Zhao}}, \bibinfo {author} {\bibfnamefont {M.}~\bibnamefont {Tang}},\ and\
  \bibinfo {author} {\bibfnamefont {Z.}~\bibnamefont {Xu}},\ }\bibfield
  {title} {\bibinfo {title} {{Gravitational Lensing Effects and Observational
  Behaviors of the Rotating Short-Hairy Black Hole}},\ }\href@noop {} {\
  (\bibinfo {year} {2024}{\natexlab{b}})},\ \Eprint
  {https://arxiv.org/abs/2408.01205} {arXiv:2408.01205 [gr-qc]} \BibitemShut
  {NoStop}%
\bibitem [{\citenamefont {Shakura}\ and\ \citenamefont
  {Sunyaev}(1973)}]{Shakura:1972te}%
  \BibitemOpen
  \bibfield  {author} {\bibinfo {author} {\bibfnamefont {N.~I.}\ \bibnamefont
  {Shakura}}\ and\ \bibinfo {author} {\bibfnamefont {R.~A.}\ \bibnamefont
  {Sunyaev}},\ }\bibfield  {title} {\bibinfo {title} {{Black holes in binary
  systems. Observational appearance}},\ }\href@noop {} {\bibfield  {journal}
  {\bibinfo  {journal} {Astron. Astrophys.}\ }\textbf {\bibinfo {volume}
  {24}},\ \bibinfo {pages} {337} (\bibinfo {year} {1973})}\BibitemShut
  {NoStop}%
\bibitem [{\citenamefont {Page}\ and\ \citenamefont
  {Thorne}(1974)}]{Page:1974he}%
  \BibitemOpen
  \bibfield  {author} {\bibinfo {author} {\bibfnamefont {D.~N.}\ \bibnamefont
  {Page}}\ and\ \bibinfo {author} {\bibfnamefont {K.~S.}\ \bibnamefont
  {Thorne}},\ }\bibfield  {title} {\bibinfo {title} {{Disk-Accretion onto a
  Black Hole. Time-Averaged Structure of Accretion Disk}},\ }\href
  {https://doi.org/10.1086/152990} {\bibfield  {journal} {\bibinfo  {journal}
  {Astrophys. J.}\ }\textbf {\bibinfo {volume} {191}},\ \bibinfo {pages} {499}
  (\bibinfo {year} {1974})}\BibitemShut {NoStop}%
\bibitem [{\citenamefont {Thorne}(1974)}]{Thorne:1974ve}%
  \BibitemOpen
  \bibfield  {author} {\bibinfo {author} {\bibfnamefont {K.~S.}\ \bibnamefont
  {Thorne}},\ }\bibfield  {title} {\bibinfo {title} {{Disk accretion onto a
  black hole. 2. Evolution of the hole.}},\ }\href
  {https://doi.org/10.1086/152991} {\bibfield  {journal} {\bibinfo  {journal}
  {Astrophys. J.}\ }\textbf {\bibinfo {volume} {191}},\ \bibinfo {pages} {507}
  (\bibinfo {year} {1974})}\BibitemShut {NoStop}%
\bibitem [{\citenamefont {Bambi}\ \emph {et~al.}(2014)\citenamefont {Bambi},
  \citenamefont {Malafarina},\ and\ \citenamefont {Tsukamoto}}]{Bambi:2014koa}%
  \BibitemOpen
  \bibfield  {author} {\bibinfo {author} {\bibfnamefont {C.}~\bibnamefont
  {Bambi}}, \bibinfo {author} {\bibfnamefont {D.}~\bibnamefont {Malafarina}},\
  and\ \bibinfo {author} {\bibfnamefont {N.}~\bibnamefont {Tsukamoto}},\
  }\bibfield  {title} {\bibinfo {title} {{Note on the effect of a massive
  accretion disk in the measurements of black hole spins}},\ }\href
  {https://doi.org/10.1103/PhysRevD.89.127302} {\bibfield  {journal} {\bibinfo
  {journal} {Phys. Rev. D}\ }\textbf {\bibinfo {volume} {89}},\ \bibinfo
  {pages} {127302} (\bibinfo {year} {2014})},\ \Eprint
  {https://arxiv.org/abs/1406.2181} {arXiv:1406.2181 [gr-qc]} \BibitemShut
  {NoStop}%
\bibitem [{\citenamefont {Kong}\ \emph {et~al.}(2014)\citenamefont {Kong},
  \citenamefont {Li},\ and\ \citenamefont {Bambi}}]{Kong:2014wha}%
  \BibitemOpen
  \bibfield  {author} {\bibinfo {author} {\bibfnamefont {L.}~\bibnamefont
  {Kong}}, \bibinfo {author} {\bibfnamefont {Z.}~\bibnamefont {Li}},\ and\
  \bibinfo {author} {\bibfnamefont {C.}~\bibnamefont {Bambi}},\ }\bibfield
  {title} {\bibinfo {title} {{Constraints on the spacetime geometry around 10
  stellar-mass black hole candidates from the disk's thermal spectrum}},\
  }\href {https://doi.org/10.1088/0004-637X/797/2/78} {\bibfield  {journal}
  {\bibinfo  {journal} {Astrophys. J.}\ }\textbf {\bibinfo {volume} {797}},\
  \bibinfo {pages} {78} (\bibinfo {year} {2014})},\ \Eprint
  {https://arxiv.org/abs/1405.1508} {arXiv:1405.1508 [gr-qc]} \BibitemShut
  {NoStop}%
\bibitem [{\citenamefont {Pun}\ \emph {et~al.}(2008{\natexlab{a}})\citenamefont
  {Pun}, \citenamefont {Kovacs},\ and\ \citenamefont {Harko}}]{Pun:2008ua}%
  \BibitemOpen
  \bibfield  {author} {\bibinfo {author} {\bibfnamefont {C.~S.~J.}\
  \bibnamefont {Pun}}, \bibinfo {author} {\bibfnamefont {Z.}~\bibnamefont
  {Kovacs}},\ and\ \bibinfo {author} {\bibfnamefont {T.}~\bibnamefont
  {Harko}},\ }\bibfield  {title} {\bibinfo {title} {{Thin accretion disks onto
  brane world black holes}},\ }\href
  {https://doi.org/10.1103/PhysRevD.78.084015} {\bibfield  {journal} {\bibinfo
  {journal} {Phys. Rev. D}\ }\textbf {\bibinfo {volume} {78}},\ \bibinfo
  {pages} {084015} (\bibinfo {year} {2008}{\natexlab{a}})},\ \Eprint
  {https://arxiv.org/abs/0809.1284} {arXiv:0809.1284 [gr-qc]} \BibitemShut
  {NoStop}%
\bibitem [{\citenamefont {Harko}\ \emph {et~al.}(2009)\citenamefont {Harko},
  \citenamefont {Kovacs},\ and\ \citenamefont {Lobo}}]{Harko:2009rp}%
  \BibitemOpen
  \bibfield  {author} {\bibinfo {author} {\bibfnamefont {T.}~\bibnamefont
  {Harko}}, \bibinfo {author} {\bibfnamefont {Z.}~\bibnamefont {Kovacs}},\ and\
  \bibinfo {author} {\bibfnamefont {F.~S.~N.}\ \bibnamefont {Lobo}},\
  }\bibfield  {title} {\bibinfo {title} {{Testing Ho\v{r}ava-Lifshitz gravity
  using thin accretion disk properties}},\ }\href
  {https://doi.org/10.1103/PhysRevD.80.044021} {\bibfield  {journal} {\bibinfo
  {journal} {Phys. Rev. D}\ }\textbf {\bibinfo {volume} {80}},\ \bibinfo
  {pages} {044021} (\bibinfo {year} {2009})},\ \Eprint
  {https://arxiv.org/abs/0907.1449} {arXiv:0907.1449 [gr-qc]} \BibitemShut
  {NoStop}%
\bibitem [{\citenamefont {Harko}\ \emph {et~al.}(2010)\citenamefont {Harko},
  \citenamefont {Kovacs},\ and\ \citenamefont {Lobo}}]{Harko:2009kj}%
  \BibitemOpen
  \bibfield  {author} {\bibinfo {author} {\bibfnamefont {T.}~\bibnamefont
  {Harko}}, \bibinfo {author} {\bibfnamefont {Z.}~\bibnamefont {Kovacs}},\ and\
  \bibinfo {author} {\bibfnamefont {F.~S.~N.}\ \bibnamefont {Lobo}},\
  }\bibfield  {title} {\bibinfo {title} {{Thin accretion disk signatures in
  dynamical Chern-Simons modified gravity}},\ }\href
  {https://doi.org/10.1088/0264-9381/27/10/105010} {\bibfield  {journal}
  {\bibinfo  {journal} {Class. Quant. Grav.}\ }\textbf {\bibinfo {volume}
  {27}},\ \bibinfo {pages} {105010} (\bibinfo {year} {2010})},\ \Eprint
  {https://arxiv.org/abs/0909.1267} {arXiv:0909.1267 [gr-qc]} \BibitemShut
  {NoStop}%
\bibitem [{\citenamefont {Harko}\ \emph {et~al.}(2011)\citenamefont {Harko},
  \citenamefont {Kovacs},\ and\ \citenamefont {Lobo}}]{Harko:2010ua}%
  \BibitemOpen
  \bibfield  {author} {\bibinfo {author} {\bibfnamefont {T.}~\bibnamefont
  {Harko}}, \bibinfo {author} {\bibfnamefont {Z.}~\bibnamefont {Kovacs}},\ and\
  \bibinfo {author} {\bibfnamefont {F.~S.~N.}\ \bibnamefont {Lobo}},\
  }\bibfield  {title} {\bibinfo {title} {{Thin accretion disk signatures of
  slowly rotating black holes in Ho\v{r}ava gravity}},\ }\href
  {https://doi.org/10.1088/0264-9381/28/16/165001} {\bibfield  {journal}
  {\bibinfo  {journal} {Class. Quant. Grav.}\ }\textbf {\bibinfo {volume}
  {28}},\ \bibinfo {pages} {165001} (\bibinfo {year} {2011})},\ \Eprint
  {https://arxiv.org/abs/1009.1958} {arXiv:1009.1958 [gr-qc]} \BibitemShut
  {NoStop}%
\bibitem [{\citenamefont {Chakraborty}(2015)}]{Chakraborty:2014eha}%
  \BibitemOpen
  \bibfield  {author} {\bibinfo {author} {\bibfnamefont {S.}~\bibnamefont
  {Chakraborty}},\ }\bibfield  {title} {\bibinfo {title} {{Equilibrium
  configuration of perfect fluid orbiting around black holes in some classes of
  alternative gravity theories}},\ }\href
  {https://doi.org/10.1088/0264-9381/32/7/075007} {\bibfield  {journal}
  {\bibinfo  {journal} {Class. Quant. Grav.}\ }\textbf {\bibinfo {volume}
  {32}},\ \bibinfo {pages} {075007} (\bibinfo {year} {2015})},\ \Eprint
  {https://arxiv.org/abs/1406.0417} {arXiv:1406.0417 [gr-qc]} \BibitemShut
  {NoStop}%
\bibitem [{\citenamefont {Pun}\ \emph {et~al.}(2008{\natexlab{b}})\citenamefont
  {Pun}, \citenamefont {Kovacs},\ and\ \citenamefont {Harko}}]{Pun:2008ae}%
  \BibitemOpen
  \bibfield  {author} {\bibinfo {author} {\bibfnamefont {C.~S.~J.}\
  \bibnamefont {Pun}}, \bibinfo {author} {\bibfnamefont {Z.}~\bibnamefont
  {Kovacs}},\ and\ \bibinfo {author} {\bibfnamefont {T.}~\bibnamefont
  {Harko}},\ }\bibfield  {title} {\bibinfo {title} {{Thin accretion disks in
  f(R) modified gravity models}},\ }\href
  {https://doi.org/10.1103/PhysRevD.78.024043} {\bibfield  {journal} {\bibinfo
  {journal} {Phys. Rev. D}\ }\textbf {\bibinfo {volume} {78}},\ \bibinfo
  {pages} {024043} (\bibinfo {year} {2008}{\natexlab{b}})},\ \Eprint
  {https://arxiv.org/abs/0806.0679} {arXiv:0806.0679 [gr-qc]} \BibitemShut
  {NoStop}%
\bibitem [{\citenamefont {Li}\ \emph {et~al.}(2005)\citenamefont {Li},
  \citenamefont {Zimmerman}, \citenamefont {Narayan},\ and\ \citenamefont
  {McClintock}}]{Li:2004aq}%
  \BibitemOpen
  \bibfield  {author} {\bibinfo {author} {\bibfnamefont {L.-X.}\ \bibnamefont
  {Li}}, \bibinfo {author} {\bibfnamefont {E.~R.}\ \bibnamefont {Zimmerman}},
  \bibinfo {author} {\bibfnamefont {R.}~\bibnamefont {Narayan}},\ and\ \bibinfo
  {author} {\bibfnamefont {J.~E.}\ \bibnamefont {McClintock}},\ }\bibfield
  {title} {\bibinfo {title} {{Multi-temperature blackbody spectrum of a thin
  accretion disk around a Kerr black hole: Model computations and comparison
  with observations}},\ }\href {https://doi.org/10.1086/428089} {\bibfield
  {journal} {\bibinfo  {journal} {Astrophys. J. Suppl.}\ }\textbf {\bibinfo
  {volume} {157}},\ \bibinfo {pages} {335} (\bibinfo {year} {2005})},\ \Eprint
  {https://arxiv.org/abs/astro-ph/0411583} {arXiv:astro-ph/0411583}
  \BibitemShut {NoStop}%
\bibitem [{\citenamefont {Narayan}\ and\ \citenamefont
  {Popham}(1993)}]{Narayan:1993bd}%
  \BibitemOpen
  \bibfield  {author} {\bibinfo {author} {\bibfnamefont {R.}~\bibnamefont
  {Narayan}}\ and\ \bibinfo {author} {\bibfnamefont {R.}~\bibnamefont
  {Popham}},\ }\bibfield  {title} {\bibinfo {title} {{Hard x-rays from
  accretion disk boundary layers}},\ }\href {https://doi.org/10.1038/362820a0}
  {\bibfield  {journal} {\bibinfo  {journal} {Nature}\ }\textbf {\bibinfo
  {volume} {362}},\ \bibinfo {pages} {820} (\bibinfo {year}
  {1993})}\BibitemShut {NoStop}%
\bibitem [{\citenamefont {Bambi}(2017{\natexlab{a}})}]{Bambi:2017khi}%
  \BibitemOpen
  \bibfield  {author} {\bibinfo {author} {\bibfnamefont {C.}~\bibnamefont
  {Bambi}},\ }\href {https://doi.org/10.1007/978-981-10-4524-0} {\emph
  {\bibinfo {title} {{Black Holes: A Laboratory for Testing Strong Gravity}}}}\
  (\bibinfo  {publisher} {Springer},\ \bibinfo {year} {2017})\BibitemShut
  {NoStop}%
\bibitem [{\citenamefont {Heydari-Fard}(2010)}]{Heydari-Fard:2010agr}%
  \BibitemOpen
  \bibfield  {author} {\bibinfo {author} {\bibfnamefont {M.}~\bibnamefont
  {Heydari-Fard}},\ }\bibfield  {title} {\bibinfo {title} {{Black hole
  accretion disks in brane gravity via a confining potential}},\ }\href
  {https://doi.org/10.1088/0264-9381/27/23/235004} {\bibfield  {journal}
  {\bibinfo  {journal} {Class. Quant. Grav.}\ }\textbf {\bibinfo {volume}
  {27}},\ \bibinfo {pages} {235004} (\bibinfo {year} {2010})}\BibitemShut
  {NoStop}%
\bibitem [{\citenamefont {Dyadina}\ and\ \citenamefont
  {Avdeev}(2024)}]{Dyadina:2010agr}%
  \BibitemOpen
  \bibfield  {author} {\bibinfo {author} {\bibfnamefont {P.}~\bibnamefont
  {Dyadina}}\ and\ \bibinfo {author} {\bibfnamefont {N.}~\bibnamefont
  {Avdeev}},\ }\bibfield  {title} {\bibinfo {title} {{Thin accretion disk
  signatures in hybrid metric-Palatini gravity}},\ }\href
  {https://doi.org/10.1140/epjc/s10052-024-12465-7} {\bibfield  {journal}
  {\bibinfo  {journal} {Eur. Phys. J. C}\ }\textbf {\bibinfo {volume} {84}},\
  \bibinfo {pages} {103} (\bibinfo {year} {2024})},\ \Eprint
  {https://arxiv.org/abs/2308.02375} {arXiv:2308.02375 [gr-qc]} \BibitemShut
  {NoStop}%
\bibitem [{\citenamefont {Heydari-Fard}\ \emph {et~al.}(2024)\citenamefont
  {Heydari-Fard}, \citenamefont {Heydari-Fard},\ and\ \citenamefont
  {Riazi}}]{Heydari-Fard:2023kgf}%
  \BibitemOpen
  \bibfield  {author} {\bibinfo {author} {\bibfnamefont {M.}~\bibnamefont
  {Heydari-Fard}}, \bibinfo {author} {\bibfnamefont {M.}~\bibnamefont
  {Heydari-Fard}},\ and\ \bibinfo {author} {\bibfnamefont {N.}~\bibnamefont
  {Riazi}},\ }\bibfield  {title} {\bibinfo {title} {{Thin accretion disk images
  of rotating hairy Horndeski black holes}},\ }\href
  {https://doi.org/10.1007/s10509-024-04359-7} {\bibfield  {journal} {\bibinfo
  {journal} {Astrophys. Space Sci.}\ }\textbf {\bibinfo {volume} {369}},\
  \bibinfo {pages} {96} (\bibinfo {year} {2024})},\ \Eprint
  {https://arxiv.org/abs/2311.12393} {arXiv:2311.12393 [gr-qc]} \BibitemShut
  {NoStop}%
\bibitem [{\citenamefont {Donmez}(2024)}]{Donmez:2024lfi}%
  \BibitemOpen
  \bibfield  {author} {\bibinfo {author} {\bibfnamefont {O.}~\bibnamefont
  {Donmez}},\ }\bibfield  {title} {\bibinfo {title} {{Bondi-Hoyle-Lyttleton
  accretion around the rotating hairy Horndeski black hole}},\ }\href
  {https://doi.org/10.1088/1475-7516/2024/09/006} {\bibfield  {journal}
  {\bibinfo  {journal} {JCAP}\ }\textbf {\bibinfo {volume} {09}},\ \bibinfo
  {pages} {006}},\ \Eprint {https://arxiv.org/abs/2402.16707} {arXiv:2402.16707
  [astro-ph.HE]} \BibitemShut {NoStop}%
\bibitem [{\citenamefont {Boshkayev}\ \emph {et~al.}(2024)\citenamefont
  {Boshkayev}, \citenamefont {Konysbayev}, \citenamefont {Kurmanov},
  \citenamefont {Luongo}, \citenamefont {Muccino}, \citenamefont {Taukenova},\
  and\ \citenamefont {Urazalina}}]{Boshkayev:2023fft}%
  \BibitemOpen
  \bibfield  {author} {\bibinfo {author} {\bibfnamefont {K.}~\bibnamefont
  {Boshkayev}}, \bibinfo {author} {\bibfnamefont {T.}~\bibnamefont
  {Konysbayev}}, \bibinfo {author} {\bibfnamefont {Y.}~\bibnamefont
  {Kurmanov}}, \bibinfo {author} {\bibfnamefont {O.}~\bibnamefont {Luongo}},
  \bibinfo {author} {\bibfnamefont {M.}~\bibnamefont {Muccino}}, \bibinfo
  {author} {\bibfnamefont {A.}~\bibnamefont {Taukenova}},\ and\ \bibinfo
  {author} {\bibfnamefont {A.}~\bibnamefont {Urazalina}},\ }\bibfield  {title}
  {\bibinfo {title} {{Luminosity of accretion disks around rotating regular
  black holes}},\ }\href {https://doi.org/10.1140/epjc/s10052-024-12446-w}
  {\bibfield  {journal} {\bibinfo  {journal} {Eur. Phys. J. C}\ }\textbf
  {\bibinfo {volume} {84}},\ \bibinfo {pages} {230} (\bibinfo {year} {2024})},\
  \Eprint {https://arxiv.org/abs/2307.15003} {arXiv:2307.15003 [gr-qc]}
  \BibitemShut {NoStop}%
\bibitem [{\citenamefont {P\'erez}\ \emph {et~al.}(2017)\citenamefont
  {P\'erez}, \citenamefont {Lopez~Armengol},\ and\ \citenamefont
  {Romero}}]{Perez:2017spz}%
  \BibitemOpen
  \bibfield  {author} {\bibinfo {author} {\bibfnamefont {D.}~\bibnamefont
  {P\'erez}}, \bibinfo {author} {\bibfnamefont {F.~G.}\ \bibnamefont
  {Lopez~Armengol}},\ and\ \bibinfo {author} {\bibfnamefont {G.~E.}\
  \bibnamefont {Romero}},\ }\bibfield  {title} {\bibinfo {title} {{Accretion
  disks around black holes in Scalar-Tensor-Vector Gravity}},\ }\href
  {https://doi.org/10.1103/PhysRevD.95.104047} {\bibfield  {journal} {\bibinfo
  {journal} {Phys. Rev. D}\ }\textbf {\bibinfo {volume} {95}},\ \bibinfo
  {pages} {104047} (\bibinfo {year} {2017})},\ \Eprint
  {https://arxiv.org/abs/1705.02713} {arXiv:1705.02713 [astro-ph.HE]}
  \BibitemShut {NoStop}%
\bibitem [{\citenamefont {Karimov}\ \emph {et~al.}(2018)\citenamefont
  {Karimov}, \citenamefont {Izmailov}, \citenamefont {Bhattacharya},\ and\
  \citenamefont {Nandi}}]{Karimov:2018whx}%
  \BibitemOpen
  \bibfield  {author} {\bibinfo {author} {\bibfnamefont {R.~K.}\ \bibnamefont
  {Karimov}}, \bibinfo {author} {\bibfnamefont {R.~N.}\ \bibnamefont
  {Izmailov}}, \bibinfo {author} {\bibfnamefont {A.}~\bibnamefont
  {Bhattacharya}},\ and\ \bibinfo {author} {\bibfnamefont {K.~K.}\ \bibnamefont
  {Nandi}},\ }\bibfield  {title} {\bibinfo {title} {{Accretion disks around the
  Gibbons–Maeda–Garfinkle–Horowitz–Strominger charged black holes}},\
  }\href {https://doi.org/10.1140/epjc/s10052-018-6270-6} {\bibfield  {journal}
  {\bibinfo  {journal} {Eur. Phys. J. C}\ }\textbf {\bibinfo {volume} {78}},\
  \bibinfo {pages} {788} (\bibinfo {year} {2018})},\ \Eprint
  {https://arxiv.org/abs/2002.00589} {arXiv:2002.00589 [gr-qc]} \BibitemShut
  {NoStop}%
\bibitem [{\citenamefont {Jiang}\ and\ \citenamefont
  {Wang}(2024)}]{Jiang:2024njc}%
  \BibitemOpen
  \bibfield  {author} {\bibinfo {author} {\bibfnamefont {Y.-H.}\ \bibnamefont
  {Jiang}}\ and\ \bibinfo {author} {\bibfnamefont {T.}~\bibnamefont {Wang}},\
  }\bibfield  {title} {\bibinfo {title} {{Accretion disks around magnetically
  charged black holes in string theory with an Euler-Heisenberg correction}},\
  }\href {https://doi.org/10.1103/PhysRevD.110.103009} {\bibfield  {journal}
  {\bibinfo  {journal} {Phys. Rev. D}\ }\textbf {\bibinfo {volume} {110}},\
  \bibinfo {pages} {103009} (\bibinfo {year} {2024})},\ \Eprint
  {https://arxiv.org/abs/2408.10150} {arXiv:2408.10150 [gr-qc]} \BibitemShut
  {NoStop}%
\bibitem [{\citenamefont {Feng}\ \emph {et~al.}(2025)\citenamefont {Feng},
  \citenamefont {Yang},\ and\ \citenamefont {Chen}}]{Feng:2024iqj}%
  \BibitemOpen
  \bibfield  {author} {\bibinfo {author} {\bibfnamefont {H.}~\bibnamefont
  {Feng}}, \bibinfo {author} {\bibfnamefont {R.-J.}\ \bibnamefont {Yang}},\
  and\ \bibinfo {author} {\bibfnamefont {W.-Q.}\ \bibnamefont {Chen}},\
  }\bibfield  {title} {\bibinfo {title} {{Thin accretion disk and shadow of
  Kerr\textendash{}Sen black hole in
  Einstein\textendash{}Maxwell-dilaton\textendash{}axion gravity}},\ }\href
  {https://doi.org/10.1016/j.astropartphys.2024.103075} {\bibfield  {journal}
  {\bibinfo  {journal} {Astropart. Phys.}\ }\textbf {\bibinfo {volume} {166}},\
  \bibinfo {pages} {103075} (\bibinfo {year} {2025})},\ \Eprint
  {https://arxiv.org/abs/2403.18541} {arXiv:2403.18541 [gr-qc]} \BibitemShut
  {NoStop}%
\bibitem [{\citenamefont {Heydari-Fard}\ \emph {et~al.}(2023)\citenamefont
  {Heydari-Fard}, \citenamefont {Honarvar},\ and\ \citenamefont
  {Heydari-Fard}}]{Heydari-Fard:2022xhr}%
  \BibitemOpen
  \bibfield  {author} {\bibinfo {author} {\bibfnamefont {M.}~\bibnamefont
  {Heydari-Fard}}, \bibinfo {author} {\bibfnamefont {S.~G.}\ \bibnamefont
  {Honarvar}},\ and\ \bibinfo {author} {\bibfnamefont {M.}~\bibnamefont
  {Heydari-Fard}},\ }\bibfield  {title} {\bibinfo {title} {{Thin accretion disc
  luminosity and its image around rotating black holes in perfect fluid dark
  matter}},\ }\href {https://doi.org/10.1093/mnras/stad558} {\bibfield
  {journal} {\bibinfo  {journal} {Mon. Not. Roy. Astron. Soc.}\ }\textbf
  {\bibinfo {volume} {521}},\ \bibinfo {pages} {708} (\bibinfo {year}
  {2023})},\ \Eprint {https://arxiv.org/abs/2210.04173} {arXiv:2210.04173
  [gr-qc]} \BibitemShut {NoStop}%
\bibitem [{\citenamefont {Heydari-Fard}\ \emph {et~al.}(2021)\citenamefont
  {Heydari-Fard}, \citenamefont {Heydari-Fard},\ and\ \citenamefont
  {Sepangi}}]{Heydari-Fard:2021ljh}%
  \BibitemOpen
  \bibfield  {author} {\bibinfo {author} {\bibfnamefont {M.}~\bibnamefont
  {Heydari-Fard}}, \bibinfo {author} {\bibfnamefont {M.}~\bibnamefont
  {Heydari-Fard}},\ and\ \bibinfo {author} {\bibfnamefont {H.~R.}\ \bibnamefont
  {Sepangi}},\ }\bibfield  {title} {\bibinfo {title} {{Thin accretion disks
  around rotating black holes in 4$D$
  Einstein\textendash{}Gauss\textendash{}Bonnet gravity}},\ }\href
  {https://doi.org/10.1140/epjc/s10052-021-09266-7} {\bibfield  {journal}
  {\bibinfo  {journal} {Eur. Phys. J. C}\ }\textbf {\bibinfo {volume} {81}},\
  \bibinfo {pages} {473} (\bibinfo {year} {2021})},\ \Eprint
  {https://arxiv.org/abs/2105.09192} {arXiv:2105.09192 [gr-qc]} \BibitemShut
  {NoStop}%
\bibitem [{\citenamefont {Liu}\ \emph {et~al.}(2022)\citenamefont {Liu},
  \citenamefont {Yang}, \citenamefont {Wu},\ and\ \citenamefont
  {Zhu}}]{Liu:2021yev}%
  \BibitemOpen
  \bibfield  {author} {\bibinfo {author} {\bibfnamefont {C.}~\bibnamefont
  {Liu}}, \bibinfo {author} {\bibfnamefont {S.}~\bibnamefont {Yang}}, \bibinfo
  {author} {\bibfnamefont {Q.}~\bibnamefont {Wu}},\ and\ \bibinfo {author}
  {\bibfnamefont {T.}~\bibnamefont {Zhu}},\ }\bibfield  {title} {\bibinfo
  {title} {{Thin accretion disk onto slowly rotating black holes in
  Einstein-\AE{}ther theory}},\ }\href
  {https://doi.org/10.1088/1475-7516/2022/02/034} {\bibfield  {journal}
  {\bibinfo  {journal} {JCAP}\ }\textbf {\bibinfo {volume} {02}}\bibfield
  {number} {\bibinfo  {number} { (02)},\ \bibinfo {pages} {034}},\ }\Eprint
  {https://arxiv.org/abs/2107.04811} {arXiv:2107.04811 [gr-qc]} \BibitemShut
  {NoStop}%
\bibitem [{\citenamefont {Heydari-Fard}\ and\ \citenamefont
  {Sepangi}(2021)}]{Heydari-Fard:2020iiu}%
  \BibitemOpen
  \bibfield  {author} {\bibinfo {author} {\bibfnamefont {M.}~\bibnamefont
  {Heydari-Fard}}\ and\ \bibinfo {author} {\bibfnamefont {H.~R.}\ \bibnamefont
  {Sepangi}},\ }\bibfield  {title} {\bibinfo {title} {{Thin accretion disk
  signatures of scalarized black holes in Einstein-scalar-Gauss-Bonnet
  gravity}},\ }\href {https://doi.org/10.1016/j.physletb.2021.136276}
  {\bibfield  {journal} {\bibinfo  {journal} {Phys. Lett. B}\ }\textbf
  {\bibinfo {volume} {816}},\ \bibinfo {pages} {136276} (\bibinfo {year}
  {2021})},\ \Eprint {https://arxiv.org/abs/2009.13748} {arXiv:2009.13748
  [gr-qc]} \BibitemShut {NoStop}%
\bibitem [{\citenamefont {Heydari-Fard}\ \emph {et~al.}(2020)\citenamefont
  {Heydari-Fard}, \citenamefont {Heydari-Fard},\ and\ \citenamefont
  {Sepangi}}]{Heydari-Fard:2020ugv}%
  \BibitemOpen
  \bibfield  {author} {\bibinfo {author} {\bibfnamefont {M.}~\bibnamefont
  {Heydari-Fard}}, \bibinfo {author} {\bibfnamefont {M.}~\bibnamefont
  {Heydari-Fard}},\ and\ \bibinfo {author} {\bibfnamefont {H.~R.}\ \bibnamefont
  {Sepangi}},\ }\bibfield  {title} {\bibinfo {title} {{Thin accretion disks and
  charged rotating dilaton black holes}},\ }\href
  {https://doi.org/10.1140/epjc/s10052-020-7911-0} {\bibfield  {journal}
  {\bibinfo  {journal} {Eur. Phys. J. C}\ }\textbf {\bibinfo {volume} {80}},\
  \bibinfo {pages} {351} (\bibinfo {year} {2020})},\ \Eprint
  {https://arxiv.org/abs/2004.05552} {arXiv:2004.05552 [gr-qc]} \BibitemShut
  {NoStop}%
\bibitem [{\citenamefont {Lewandowski}\ \emph {et~al.}(2023)\citenamefont
  {Lewandowski}, \citenamefont {Ma}, \citenamefont {Yang},\ and\ \citenamefont
  {Zhang}}]{Lewandowski:2022zce}%
  \BibitemOpen
  \bibfield  {author} {\bibinfo {author} {\bibfnamefont {J.}~\bibnamefont
  {Lewandowski}}, \bibinfo {author} {\bibfnamefont {Y.}~\bibnamefont {Ma}},
  \bibinfo {author} {\bibfnamefont {J.}~\bibnamefont {Yang}},\ and\ \bibinfo
  {author} {\bibfnamefont {C.}~\bibnamefont {Zhang}},\ }\bibfield  {title}
  {\bibinfo {title} {{Quantum Oppenheimer-Snyder and Swiss Cheese Models}},\
  }\href {https://doi.org/10.1103/PhysRevLett.130.101501} {\bibfield  {journal}
  {\bibinfo  {journal} {Phys. Rev. Lett.}\ }\textbf {\bibinfo {volume} {130}},\
  \bibinfo {pages} {101501} (\bibinfo {year} {2023})},\ \Eprint
  {https://arxiv.org/abs/2210.02253} {arXiv:2210.02253 [gr-qc]} \BibitemShut
  {NoStop}%
\bibitem [{\citenamefont {Collodel}\ \emph {et~al.}(2021)\citenamefont
  {Collodel}, \citenamefont {Doneva},\ and\ \citenamefont
  {Yazadjiev}}]{Collodel:2021gxu}%
  \BibitemOpen
  \bibfield  {author} {\bibinfo {author} {\bibfnamefont {L.~G.}\ \bibnamefont
  {Collodel}}, \bibinfo {author} {\bibfnamefont {D.~D.}\ \bibnamefont
  {Doneva}},\ and\ \bibinfo {author} {\bibfnamefont {S.~S.}\ \bibnamefont
  {Yazadjiev}},\ }\bibfield  {title} {\bibinfo {title} {{Circular Orbit
  Structure and Thin Accretion Disks around Kerr Black Holes with Scalar
  Hair}},\ }\href {https://doi.org/10.3847/1538-4357/abe305} {\bibfield
  {journal} {\bibinfo  {journal} {Astrophys. J.}\ }\textbf {\bibinfo {volume}
  {910}},\ \bibinfo {pages} {52} (\bibinfo {year} {2021})},\ \Eprint
  {https://arxiv.org/abs/2101.05073} {arXiv:2101.05073 [astro-ph.HE]}
  \BibitemShut {NoStop}%
\bibitem [{\citenamefont {Wu}\ \emph {et~al.}(2024)\citenamefont {Wu},
  \citenamefont {Feng},\ and\ \citenamefont {Chen}}]{Wu:2024sng}%
  \BibitemOpen
  \bibfield  {author} {\bibinfo {author} {\bibfnamefont {Y.}~\bibnamefont
  {Wu}}, \bibinfo {author} {\bibfnamefont {H.}~\bibnamefont {Feng}},\ and\
  \bibinfo {author} {\bibfnamefont {W.-Q.}\ \bibnamefont {Chen}},\ }\bibfield
  {title} {\bibinfo {title} {{Thin accretion disk around black hole in
  Einstein\textendash{}Maxwell-scalar theory}},\ }\href
  {https://doi.org/10.1140/epjc/s10052-024-13454-6} {\bibfield  {journal}
  {\bibinfo  {journal} {Eur. Phys. J. C}\ }\textbf {\bibinfo {volume} {84}},\
  \bibinfo {pages} {1075} (\bibinfo {year} {2024})},\ \Eprint
  {https://arxiv.org/abs/2410.14113} {arXiv:2410.14113 [gr-qc]} \BibitemShut
  {NoStop}%
\bibitem [{\citenamefont {Kurmanov}\ \emph {et~al.}(2024)\citenamefont
  {Kurmanov}, \citenamefont {Boshkayev}, \citenamefont {Konysbayev},
  \citenamefont {Luongo}, \citenamefont {Saiyp}, \citenamefont {Urazalina},
  \citenamefont {Ikhsan},\ and\ \citenamefont {Suliyeva}}]{Kurmanov:2024hpn}%
  \BibitemOpen
  \bibfield  {author} {\bibinfo {author} {\bibfnamefont {Y.}~\bibnamefont
  {Kurmanov}}, \bibinfo {author} {\bibfnamefont {K.}~\bibnamefont {Boshkayev}},
  \bibinfo {author} {\bibfnamefont {T.}~\bibnamefont {Konysbayev}}, \bibinfo
  {author} {\bibfnamefont {O.}~\bibnamefont {Luongo}}, \bibinfo {author}
  {\bibfnamefont {N.}~\bibnamefont {Saiyp}}, \bibinfo {author} {\bibfnamefont
  {A.}~\bibnamefont {Urazalina}}, \bibinfo {author} {\bibfnamefont
  {G.}~\bibnamefont {Ikhsan}},\ and\ \bibinfo {author} {\bibfnamefont
  {G.}~\bibnamefont {Suliyeva}},\ }\bibfield  {title} {\bibinfo {title}
  {{Accretion disks properties around regular black hole solutions obtained
  from non-linear electrodynamics}},\ }\href
  {https://doi.org/10.1016/j.dark.2024.101566} {\bibfield  {journal} {\bibinfo
  {journal} {Phys. Dark Univ.}\ }\textbf {\bibinfo {volume} {46}},\ \bibinfo
  {pages} {101566} (\bibinfo {year} {2024})},\ \Eprint
  {https://arxiv.org/abs/2404.15437} {arXiv:2404.15437 [gr-qc]} \BibitemShut
  {NoStop}%
\bibitem [{\citenamefont {Liu}\ \emph {et~al.}(2024)\citenamefont {Liu},
  \citenamefont {He}, \citenamefont {Liu}, \citenamefont {Han},\ and\
  \citenamefont {Yang}}]{Liu:2024brf}%
  \BibitemOpen
  \bibfield  {author} {\bibinfo {author} {\bibfnamefont {A.}~\bibnamefont
  {Liu}}, \bibinfo {author} {\bibfnamefont {T.-Y.}\ \bibnamefont {He}},
  \bibinfo {author} {\bibfnamefont {M.}~\bibnamefont {Liu}}, \bibinfo {author}
  {\bibfnamefont {Z.-W.}\ \bibnamefont {Han}},\ and\ \bibinfo {author}
  {\bibfnamefont {R.-J.}\ \bibnamefont {Yang}},\ }\bibfield  {title} {\bibinfo
  {title} {{Possible signatures of higher dimension in thin accretion disk
  around brane world black hole}},\ }\href
  {https://doi.org/10.1088/1475-7516/2024/07/062} {\bibfield  {journal}
  {\bibinfo  {journal} {JCAP}\ }\textbf {\bibinfo {volume} {07}},\ \bibinfo
  {pages} {062}},\ \Eprint {https://arxiv.org/abs/2404.14131} {arXiv:2404.14131
  [gr-qc]} \BibitemShut {NoStop}%
\bibitem [{\citenamefont {Lee}\ \emph {et~al.}(2023)\citenamefont {Lee},
  \citenamefont {Hu}, \citenamefont {Guo},\ and\ \citenamefont
  {Chen}}]{Lee:2022rtg}%
  \BibitemOpen
  \bibfield  {author} {\bibinfo {author} {\bibfnamefont {T.-C.}\ \bibnamefont
  {Lee}}, \bibinfo {author} {\bibfnamefont {Z.}~\bibnamefont {Hu}}, \bibinfo
  {author} {\bibfnamefont {M.}~\bibnamefont {Guo}},\ and\ \bibinfo {author}
  {\bibfnamefont {B.}~\bibnamefont {Chen}},\ }\bibfield  {title} {\bibinfo
  {title} {{Circular orbits and polarized images of charged particles orbiting
  a Kerr black hole with a weak magnetic field}},\ }\href
  {https://doi.org/10.1103/PhysRevD.108.024008} {\bibfield  {journal} {\bibinfo
   {journal} {Phys. Rev. D}\ }\textbf {\bibinfo {volume} {108}},\ \bibinfo
  {pages} {024008} (\bibinfo {year} {2023})},\ \Eprint
  {https://arxiv.org/abs/2211.04143} {arXiv:2211.04143 [gr-qc]} \BibitemShut
  {NoStop}%
\bibitem [{\citenamefont {Asuk\"ula}\ \emph {et~al.}(2024)\citenamefont
  {Asuk\"ula}, \citenamefont {Hohmann}, \citenamefont {Karanasou},
  \citenamefont {Bahamonde}, \citenamefont {Pfeifer},\ and\ \citenamefont
  {Rosa}}]{Asukula:2023akj}%
  \BibitemOpen
  \bibfield  {author} {\bibinfo {author} {\bibfnamefont {H.}~\bibnamefont
  {Asuk\"ula}}, \bibinfo {author} {\bibfnamefont {M.}~\bibnamefont {Hohmann}},
  \bibinfo {author} {\bibfnamefont {V.}~\bibnamefont {Karanasou}}, \bibinfo
  {author} {\bibfnamefont {S.}~\bibnamefont {Bahamonde}}, \bibinfo {author}
  {\bibfnamefont {C.}~\bibnamefont {Pfeifer}},\ and\ \bibinfo {author}
  {\bibfnamefont {J.~a.~L.}\ \bibnamefont {Rosa}},\ }\bibfield  {title}
  {\bibinfo {title} {{Spherically symmetric vacuum solutions in one-parameter
  new general relativity and their phenomenology}},\ }\href
  {https://doi.org/10.1103/PhysRevD.109.064027} {\bibfield  {journal} {\bibinfo
   {journal} {Phys. Rev. D}\ }\textbf {\bibinfo {volume} {109}},\ \bibinfo
  {pages} {064027} (\bibinfo {year} {2024})},\ \Eprint
  {https://arxiv.org/abs/2311.17999} {arXiv:2311.17999 [gr-qc]} \BibitemShut
  {NoStop}%
\bibitem [{\citenamefont {Olmo}\ \emph {et~al.}(2023)\citenamefont {Olmo},
  \citenamefont {Rosa}, \citenamefont {Rubiera-Garcia},\ and\ \citenamefont
  {Saez-Chillon~Gomez}}]{Olmo:2023lil}%
  \BibitemOpen
  \bibfield  {author} {\bibinfo {author} {\bibfnamefont {G.~J.}\ \bibnamefont
  {Olmo}}, \bibinfo {author} {\bibfnamefont {J.~L.}\ \bibnamefont {Rosa}},
  \bibinfo {author} {\bibfnamefont {D.}~\bibnamefont {Rubiera-Garcia}},\ and\
  \bibinfo {author} {\bibfnamefont {D.}~\bibnamefont {Saez-Chillon~Gomez}},\
  }\bibfield  {title} {\bibinfo {title} {{Shadows and photon rings of regular
  black holes and geonic horizonless compact objects}},\ }\href
  {https://doi.org/10.1088/1361-6382/aceacd} {\bibfield  {journal} {\bibinfo
  {journal} {Class. Quant. Grav.}\ }\textbf {\bibinfo {volume} {40}},\ \bibinfo
  {pages} {174002} (\bibinfo {year} {2023})},\ \Eprint
  {https://arxiv.org/abs/2302.12064} {arXiv:2302.12064 [gr-qc]} \BibitemShut
  {NoStop}%
\bibitem [{\citenamefont {Nampalliwar}\ and\ \citenamefont
  {Bambi}()}]{Nampalliwar:2018tup}%
  \BibitemOpen
  \bibfield  {author} {\bibinfo {author} {\bibfnamefont {S.}~\bibnamefont
  {Nampalliwar}}\ and\ \bibinfo {author} {\bibfnamefont {C.}~\bibnamefont
  {Bambi}},\ }\bibfield  {title} {\bibinfo {title} {{Accreting Black Holes}},\
  }\href@noop {} {\ }\Eprint {https://arxiv.org/abs/1810.07041}
  {arXiv:1810.07041 [astro-ph.HE]} \BibitemShut {NoStop}%
\bibitem [{\citenamefont {Bhattacharyya}\ \emph {et~al.}(2001)\citenamefont
  {Bhattacharyya}, \citenamefont {Misra},\ and\ \citenamefont
  {Thampan}}]{Bhattacharyya:2000kt}%
  \BibitemOpen
  \bibfield  {author} {\bibinfo {author} {\bibfnamefont {S.}~\bibnamefont
  {Bhattacharyya}}, \bibinfo {author} {\bibfnamefont {R.}~\bibnamefont
  {Misra}},\ and\ \bibinfo {author} {\bibfnamefont {A.~V.}\ \bibnamefont
  {Thampan}},\ }\bibfield  {title} {\bibinfo {title} {{General relativistic
  spectra of accretion disks around rotating neutron stars}},\ }\href
  {https://doi.org/10.1086/319807} {\bibfield  {journal} {\bibinfo  {journal}
  {Astrophys. J.}\ }\textbf {\bibinfo {volume} {550}},\ \bibinfo {pages} {841}
  (\bibinfo {year} {2001})},\ \Eprint {https://arxiv.org/abs/astro-ph/0011519}
  {arXiv:astro-ph/0011519} \BibitemShut {NoStop}%
\bibitem [{\citenamefont {Joshi}\ \emph {et~al.}(2014)\citenamefont {Joshi},
  \citenamefont {Malafarina},\ and\ \citenamefont {Narayan}}]{Joshi:2013dva}%
  \BibitemOpen
  \bibfield  {author} {\bibinfo {author} {\bibfnamefont {P.~S.}\ \bibnamefont
  {Joshi}}, \bibinfo {author} {\bibfnamefont {D.}~\bibnamefont {Malafarina}},\
  and\ \bibinfo {author} {\bibfnamefont {R.}~\bibnamefont {Narayan}},\
  }\bibfield  {title} {\bibinfo {title} {{Distinguishing black holes from naked
  singularities through their accretion disc properties}},\ }\href
  {https://doi.org/10.1088/0264-9381/31/1/015002} {\bibfield  {journal}
  {\bibinfo  {journal} {Class. Quant. Grav.}\ }\textbf {\bibinfo {volume}
  {31}},\ \bibinfo {pages} {015002} (\bibinfo {year} {2014})},\ \Eprint
  {https://arxiv.org/abs/1304.7331} {arXiv:1304.7331 [gr-qc]} \BibitemShut
  {NoStop}%
\bibitem [{\citenamefont {Bambi}(2017{\natexlab{b}})}]{Bambi:2015kza}%
  \BibitemOpen
  \bibfield  {author} {\bibinfo {author} {\bibfnamefont {C.}~\bibnamefont
  {Bambi}},\ }\bibfield  {title} {\bibinfo {title} {{Testing black hole
  candidates with electromagnetic radiation}},\ }\href
  {https://doi.org/10.1103/RevModPhys.89.025001} {\bibfield  {journal}
  {\bibinfo  {journal} {Rev. Mod. Phys.}\ }\textbf {\bibinfo {volume} {89}},\
  \bibinfo {pages} {025001} (\bibinfo {year} {2017}{\natexlab{b}})},\ \Eprint
  {https://arxiv.org/abs/1509.03884} {arXiv:1509.03884 [gr-qc]} \BibitemShut
  {NoStop}%
\bibitem [{\citenamefont {Bambi}(2021)}]{Bambi:2020cyv}%
  \BibitemOpen
  \bibfield  {author} {\bibinfo {author} {\bibfnamefont {C.}~\bibnamefont
  {Bambi}},\ }\bibfield  {title} {\bibinfo {title} {{Testing General Relativity
  with Black Hole X-Ray Data}},\ }\href
  {https://doi.org/10.1134/S1063772921100024} {\bibfield  {journal} {\bibinfo
  {journal} {Astron. Rep.}\ }\textbf {\bibinfo {volume} {65}},\ \bibinfo
  {pages} {902} (\bibinfo {year} {2021})},\ \Eprint
  {https://arxiv.org/abs/2010.03793} {arXiv:2010.03793 [gr-qc]} \BibitemShut
  {NoStop}%
\end{thebibliography}%
\end{document}